\documentclass
[preprint,byrevtex,nofootinbib,10pt,balancelastpage,titlepage,bibnotes,pre,twocolumn]{revtex4}%
\usepackage{amsfonts}
\usepackage{amsmath}
\usepackage{amssymb}
\usepackage{graphicx}%
\providecommand{\U}[1]{\protect\rule{.1in}{.1in}}
\newtheorem{theorem}{Theorem}

\newtheorem{conclusion}[theorem]{Conclusion}

\newtheorem{definition}[theorem]{Definition}

\newtheorem{proposition}[theorem]{Proposition}
\newtheorem{remark}[theorem]{Remark}

\ifx\pdfoutput\relax\let\pdfoutput=\undefined\fi
\newcount\msipdfoutput
\ifx\pdfoutput\undefined\else
\ifcase\pdfoutput\else
\ifx\paperwidth\undefined\else
\ifdim\paperheight=0pt\relax\else\pdfpageheight\paperheight\fi
\ifdim\paperwidth=0pt\relax\else\pdfpagewidth\paperwidth\fi
\fi\fi\fi
\begin{document}
\preprint{UATP/2106}
\title{Nonequilibrium Entropy in an Extended State Space}
\author{P.D. Gujrati}
\email{pdg@uakron.edu}
\affiliation{Department of Physics, Department of Polymer Science, The University of Akron,
Akron, OH 44325}

\begin{abstract}
This chapter deals with our recent attempt to extend the notion of equilibrium
(EQ) entropy to nonequilibrium (NEQ) systems so that it can also capture
memory effects. This is done by enlarging the equilibrium state space
$\mathfrak{S}$ to $\mathfrak{S}^{\prime}$ by introducing internal variables.
These variables capture the irreversibility due to internal processes. By a
proper choice of the enlarged state space $\mathfrak{S}^{\prime}$, the entropy
becomes a state function, which shares many properties of the EQ\ entropy,
except for a nonzero irreversible entropy generation. We give both a
thermodynamic and statistical extension of the entropy and prove their
equivalence in all cases by taking an appropriate $\mathfrak{S}^{\prime}$.
This provides a general nonnegative statistical expression of the entropy for
any situation. We use the statistical formulation to prove the second law. We
give several examples to determine the required internal variables, which we
then apply to several cases of interest to calculate the entropy generation.
We also provide a possible explanation for why the entropy in the classical
continuum $1$-d Tonks gas can become negative by considering a lattice model
for which the entropy is always nonnegative.

\end{abstract}
\date{\today}
\maketitle

\section{Introduction}

\subsection{Entropy as a Primitive Concept}

What distinguishes a thermodynamic system $\Sigma$, the focus of our study
here, from a mechanical system is the concepts of the entropy $S$ and the
temperature $T$, both of which are new concepts in thermodynamics without any
mechanical analogs. To be useful, $S$ and $T$ must uniquely refer to the
thermodynamic state, simply called the state and denoted by $\mathcal{M}%
$\ here, of $\Sigma$. Being functions of the state of $\Sigma$, they must be
interrelated in some fashion so only one of them can be treated as a primitive
concept, which we take to be the entropy to describe it. Although $S$ plays
important roles in diverse fields ranging from classical thermodynamics of
Clausius
\cite{Clausius,Gibbs,Rice,Tolman,Landau,DeDonder,Prigogine71,Prigogine,deGroot,Eu0,Kuiken,Ottinger,Bedeaux,Evans,Fermi,Reif,Woods,Kestin,Boltzmann,Waldram,Balian,Callen}%
, quantum mechanics and uncertainty \cite{von Neumann,Landau-QM,Partovi},
black holes \cite{Beckenstein}, coding and computation
\cite{Schumacher,Bennet}, to information technology \cite{Wiener,Shannon}, it
does not seem to have a standard definition in all cases, even though it is
well defined under equilibrium (EQ) conditions as extensively discussed in the
literature; see, for example \cite{Gujrati-Entropy1,Gujrati-Entropy2}.
Therefore, in this chapter, we are interested in understanding this concept in
nonequilibrium (NEQ) statistical thermodynamics of irreversible processes (we
will use NEQT to denote NEQ\ thermodynamics in this chapter). This will
require an extension of the classical concept of entropy from EQ states to NEQ
states where irreversible entropy will be generated.
\begin{figure}
[ptb]
\begin{center}
\includegraphics[
height=3.4584in,
width=3.3044in
]%
{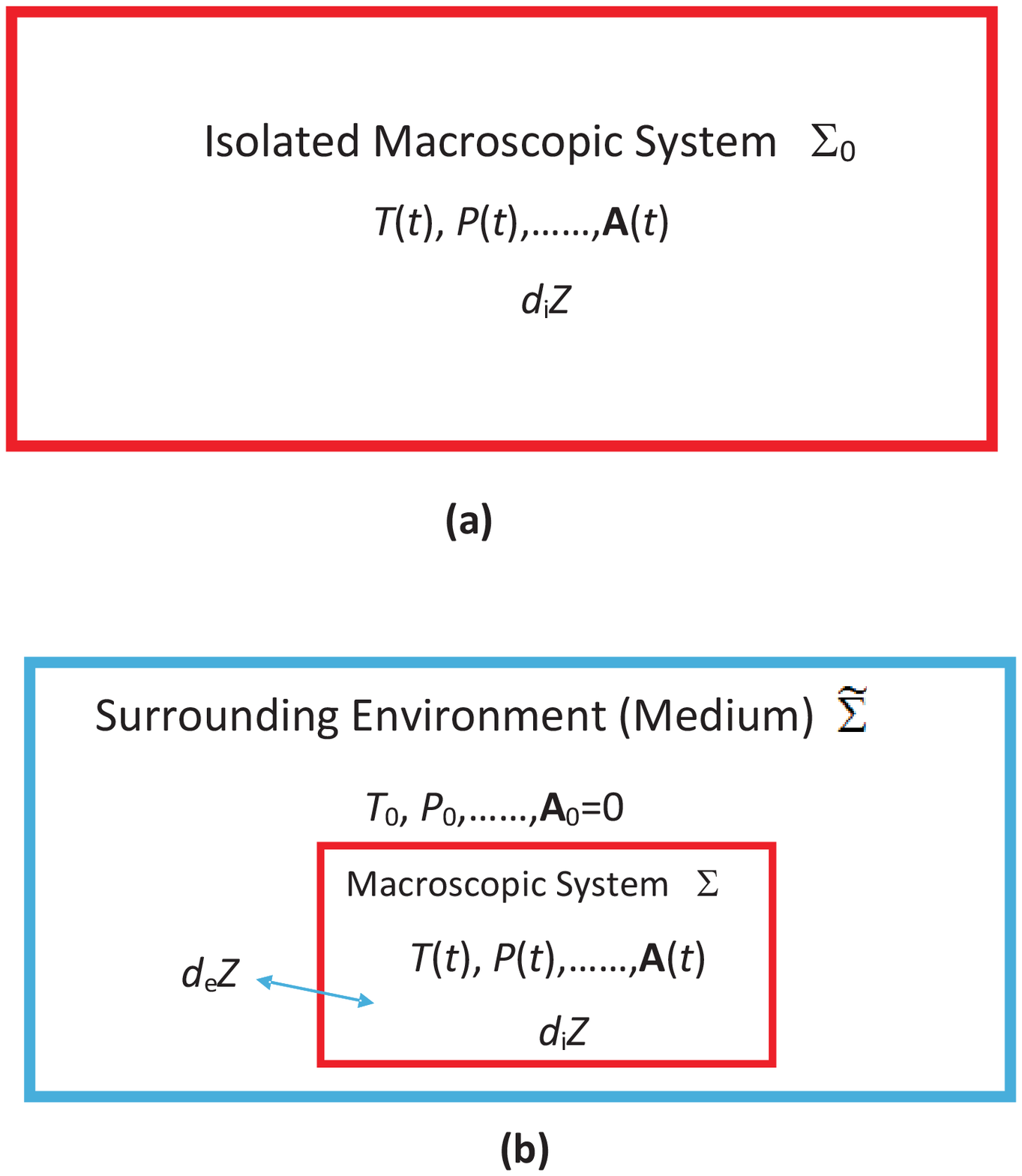}%
\caption{{}(a) An isolated nonequilibrium system $\Sigma_{0}$\ with internally
generated $d_{\text{i}}Z$ driving it towards equilibrium, during which its
SI-fields $T(t),P(t),\cdots,\mathbf{A}(t)$ continue to change to their
equilibrium values; $d_{\text{i}}Z_{k}$ denote the microanalog of
$d_{\text{i}}Z$. The sign of $d_{\text{i}}Z$ is determined by the second law.
(b) A nonequilibrium systen $\Sigma$ in a surrounding medium $\widetilde
{\Sigma}$, both forming an isolated system $\Sigma_{0}$. The macrostates of
the medium and the system are characterized by their fields $T_{0}%
,P_{0},...,\mathbf{A}_{0}=0$ and $T(t),P(t),...,\mathbf{A}(t)$, respectively,
which are different when the two are out of equilibrium. Exchange quantities
($d_{\text{e}}Z$) carry a suffix "e" and irreversibly generated quantities
($d_{\text{i}}Z$) within the system by a suffix "i" by extending the Prigogine
notation. Their sum $d_{\text{e}}Z+d_{\text{i}}Z$ is denoted by $dZ$, which is
a system-intrinsic quantity (see text). }%
\label{Fig.Sys}%
\end{center}
\end{figure}

There are two distinct approaches to understand the physics of $S$. One is the
original approach of Clausius \cite{Clausius} in classical thermodynamics,
where it appears as a primitive concept, and determines the Gibbs fundamental
relation from which follows the entire thermodynamics.\ The other approach is
the statistical one introduced by Boltzmann \cite{Boltzmann} and Gibbs
\cite{Gibbs}. The statistical extension requires dealing with the set of
\emph{microstates} $\left\{  \mathfrak{m}_{k}\right\}  ,k=1,2,\cdots$, which
we take to be countable. What we mean by a microstate is simply this: a
microstate $\mathfrak{m}$ is nothing but an instantaneous state of $\Sigma$.
If we consider an instantaneous ensemble \cite{Gibbs} of $\Sigma$ by making a
large number of its replicas at any instant $t$, all prepared under identical
conditions, then $\left\{  \mathfrak{m}_{k}\right\}  $ is the set formed from
distinct microstates at that time. Each microstate is specified by the
instantaneous values of the state variables in the corresponding replica. How
often a particular microstate $\mathfrak{m}_{k}$ appears among the replicas
determines its probability $p_{k}$. The connection with classical
thermodynamics is through the ensemble average $\left\langle Z\right\rangle $
of some quantity $Z$, which is also written simply as $Z$:
\begin{equation}
Z\mathcal{=}\left\langle Z\right\rangle \mathcal{=}%
{\textstyle\sum\nolimits_{k}}
p_{k}Z_{k}, \label{Ensemble Av}%
\end{equation}
where $Z_{k}$ is value $Z$ takes on $\mathfrak{m}_{k}$. Thus, the statistical
thermodynamics requires a probabilistic approach and deals directly with the
set $\left\{  \mathfrak{m}_{k},p_{k}\right\}  $. In contrast, the classical
thermodynamics has no association with microstates and/or their probabilities.

The system $\Sigma$ is usually embedded in a medium $\widetilde{\Sigma}$ as
shown in Fig. \ref{Fig.Sys}. Sometime, it can also be driven by inserting it
between two non-interacting media such as a rod between two heat sources. Even
though $\Sigma$\ is macroscopic in size, it is extremely small compared to the
medium $\widetilde{\Sigma}$ so it does not affect the state of $\widetilde
{\Sigma}$. We will always assume that $\widetilde{\Sigma}$ is in EQ so there
cannot be any irreversibility in it. Any irreversibility will be associated
with $\Sigma$. The collection $\Sigma_{0}=\Sigma\cup\widetilde{\Sigma}$ forms
an isolated system, which we assume to be stationary. We use a suffix $0$ to
denote all quantities pertaining to $\Sigma_{0}$, a tilde $(\widetilde{})$ for
all quantities pertaining to $\widetilde{\Sigma}$, and no suffix for all
quantities pertaining to $\Sigma$ even if it is isolated.

We will use the term "body" to refer to any of $\Sigma,\widetilde{\Sigma}$,
and $\Sigma_{0}$ and use $\Sigma_{\text{b}}$ to denote it. However, to avoid
notational complication, we will use the notation suitable for $\Sigma$ for
$\Sigma_{\text{b}}$ if no confusion would arise in the context. The states for
$\Sigma_{\text{b}}$ are determined by the set $\mathbf{X}$ of extensive
observables (the energy $E$, the volume $V$, the number of particles $N$,
etc.) specifying it$.$ Thus, we need $\mathbf{X}_{0},\widetilde{\mathbf{X}}$
and $\mathbf{X}$, respectively, for $\Sigma,\widetilde{\Sigma}$, and
$\Sigma_{0}$ for their description.

Let $\Gamma(\mathbf{X})$ be the microstate space containing $W(\mathbf{X})$
distinct microstates for $\Sigma_{\text{b}}$.\ While a temporal evolution for
$\Sigma_{\text{b}}$ is not our primary interest in this chapter, we still need
to remember the importance of temporal evolution in any thermodynamic study.
We will say that two microstates belonging to the microstate subspace
$\Gamma(\mathbf{X})$ are "connected" if one evolves from the other after some
time $\tau_{\text{c}}<\infty$. Before this time, they will be treated as
"disconnected." Let $\tau_{\text{c,max}}$ denote the maximum $\tau_{\text{c}}$
over all pairs of microstates. The space $\Gamma(\mathbf{X})$ is \emph{simply
connected} for all times longer than $\tau_{\text{c,max}}$ in that each
microstate can evolve into another microstate $\in\Gamma(\mathbf{X})$ in due
time. For $t<\tau_{\text{c,max}}$, the space $\Gamma(\mathbf{X})$ will consist
of \emph{disjoint components,} an issue that neither Boltzmann nor Gibbs has
considered to the best of our knowledge. But the issue, which we consider
later in Sect. \ref{Marker_Disjoint Space}, becomes important in considering
nonequilibrium states, especially in glasses.

\subsection{EQ Entropies}

We first discuss the EQ statistical entropy in the literature. Boltzmann
assumes \emph{equal probability} of various microstates in the simply
connected set $\Gamma(\mathbf{X})$, in which\ $\tau_{\text{eq}}=\tau
_{\text{eq}}(\mathbf{X})$ represents the equilibration time for $\Sigma
_{\text{b}}$.\ Under the equiprobable assumption, Boltzmann identifies the
entropy in terms of $W\left(  \mathbf{X}\right)  $ \cite{Planck,Landau}:
\begin{equation}
S_{\text{B}}\left(  \mathbf{X}\right)  \equiv\ln W\left(  \mathbf{X}\right)
\text{;} \label{Boltzmann_S}%
\end{equation}
we will set the Boltzmann constant to be unity so that the entropy will always
be a \emph{pure number}. The idea behind the above formula, which we will call
the \emph{Boltzmann's postulate}, implicitly appears for the first time in a
paper \cite{Boltzmann} by Boltzmann, and then appears more or less in the
above form later in his lectures \cite{Boltzmann0} where he introduces the
combinatorial approach for the first time to statistical mechanics. The
formula itself does not appear but is implied when he takes the logarithm of
the number of combinations \cite{Boltzmann0,Note-Boltzmann}, an approach that
we will adopt later in Sect. \ref{Marker_NonEq-S}. (There is another
formulation for entropy given by Boltzmann \cite{Boltzmann0,Boltzmann}, which
is also known as the Boltzmann entropy \cite{Jaynes} that we will discuss
later and that has a restricted validity; see Eq. (\ref{Boltzmann_S_1}).)
Gibbs, also using the probabilistic approach, gives the following formula for
the entropy in a \emph{canonical ensemble} \cite{Gibbs,Landau}%
\begin{equation}
S_{\text{G}}^{(\text{c})}\equiv-%
{\textstyle\sum\limits_{k}}
p_{k}^{(\text{c})}\ln p_{k}^{(\text{c})};\ \
{\textstyle\sum\limits_{k}}
p_{k}^{(\text{c})}=1 \label{Gibbs_S}%
\end{equation}
where $p_{k}^{(\text{c})}$ is the canonical ensemble probability of
$\mathfrak{m}_{k}$ of energy $E_{k}$, and the sum is over all microstates
(with other elements in $\mathbf{X}$ besides $E$ held fixed). The Gibbsian
approach assumes an ensemble at a given instant, while the Boltzmann approach
considers the evolution of a particular system in time; see for example a
recent review \cite{Gujrati-Symmetry}. In equilibrium, both entropy
expressions yield the same result. In quantum mechanics, this entropy is given
by the von Neumann entropy formulation \cite{von Neumann,Landau-QM} in terms
of the density matrix $\rho$:%
\[
S_{\text{vN}}=-Tr(\rho\ln\rho).
\]
The entropy formulation in the information theory \cite{Wiener,Shannon} has a
form that appears to be similar in form to the above Gibbs entropy even though
the temperature has no significance in the information theory. There is also
another statistical formulation of entropy, heavily used in the literature, in
terms of the phase space distribution function $f(x,t)$, which follows from
Boltzmann's celebrated H-theorem:%
\begin{equation}
S_{f}(t)=-%
{\textstyle\int}
f(x,t)\ln f(x,t)dx; \label{Sf_S}%
\end{equation}
here, $x$ denotes a point in the phase space. This quantity is not only not
dimensionless but also not the correct formulation in general
\cite{Gujrati-Entropy1,Gujrati-Entropy2}.

We now turn to Clausius' thermodynamic entropy $S$. It is oblivious to the set
$\left\{  \mathfrak{m}_{k},p_{k}\right\}  $ and deals only with the
observables of the body $\mathbf{X}=(E,V,N,\cdots)$. In equilibrium,
$S=S(\mathbf{X})~$is a \emph{state function} of $\mathbf{X}$. This functional
dependence results in the Gibbs fundamental relation
\begin{equation}
dS=\left(  \partial S/\partial\mathbf{X}\right)  \cdot d\mathbf{X.}
\label{Gibbs_Fundamental}%
\end{equation}
For a lattice model, $S$ is non-negative in accordance with the Boltzmann
definition of $S\left(  \mathbf{X}\right)  $, but is known to become negative
for a continuum model such as for an ideal gas. The latter observation implies
that such continuum models are not realistic as they violate Nernst's
postulate (the third law) and require \emph{quantum mechanics} to ensure
non-negativity of the entropy \cite{Landau}. Even the change $\Delta S$, the
heat capacity, etc. do not satisfy thermodynamic consequences of Nernst's postulate.

By invoking Nernst's postulate, which according to Planck states that the
equilibrium entropy vanishes at absolute zero, one can determine the
equilibrium entropy everywhere \emph{uniquely}. The consensus is that in EQ,
the thermodynamic entropy is not different from the above statistical
entropies due to Boltzmann and Gibbs, the exceptions being the negative
classical entropies. However, there is at present no consensus when the system
is out of equilibrium. There is also some doubt whether the nonequilibrium
thermodynamic entropy has any meaning in classical thermodynamics. We will
follow Clausius\ and take the view here that the thermodynamic entropy is a
well-defined notion even for an irreversible process going on in a body (see
\cite{Gujrati-Entropy1,Gujrati-Entropy2} for supporting arguments) for which
Clausius \cite[p. 214]{Clausius} writes
\begin{equation}
TdS>d_{\text{e}}Q \label{Clausius Inequality}%
\end{equation}
in terms of the exchange heat $d_{\text{e}}Q=T_{0}d_{\text{e}}S$ with the
medium at temperature $T_{0}$; see Fig. \ref{Fig.Sys}. The question that
arises is whether the statistical definitions above can be applied to a body
out of equilibrium. We find the answer to be affirmative. The next question
that arises is the following: Do they always give the same results?\ We will
show that under certain conditions, they give the same results. This is
important as the literature is not very clear on this issue
\cite{Lavis,Bishop,Lebowitz,Ruelle}.\ 

For an isolated system, we are not concerned with any thermostat or external
working source. As a consequence, $E_{0}$,$V_{0}$,$N_{0}$, etc. in
$\mathbf{X}_{0}$ must remain constant even if the system is out of
equilibrium. While this will simplify our discussion to some extent for
$\Sigma_{\text{b}}=\Sigma_{0}$, it will also create a problem discussed in the following

\begin{remark}
\label{Remark-IsolatedSystem} For an isolated system, all the observables in
$\mathbf{X}_{0}$ are fixed so if the entropy is a function of $\mathbf{X}_{0}$
only, it \emph{cannot} change
\cite{Gujrati-I,Gujrati-II,Gujrati-Entropy1,Gujrati-Entropy2} even if the
system is out of EQ.
\end{remark}

Thus, we need additional independent variables to ensure the law of increase
of entropy for a NEQ isolated system. There must be internal irreversible
processes as discussed in the next section, where we discuss the concept of a
nonequilibrium state, state variables and state functions. We show that the
situation requires a NEQ state $\mathcal{M}$ to be discussed in an extended
state space $\mathfrak{S}_{\mathbf{Z}}$ formed by the union $\mathbf{Z}%
=\mathbf{X\cup}\boldsymbol{\xi}$, where $\boldsymbol{\xi}$ is a set of
internal variables. In an appropriately chosen $\mathfrak{S}_{\mathbf{Z}}$,
the entropy of $\mathcal{M}$ becomes a unique state function of $\mathbf{Z}$.
Such a state $\mathcal{M}_{\text{ieq}}$ is identified as in internal EQ (IEQ)
in $\mathfrak{S}_{\mathbf{Z}}$, and shares many property of an EQ state,
except that it has irreversible entropy generation.\ In Sect.
\ref{Sec-InternalVariables}, we consider various situations to identify the
required number of internal variables to guide us to identify $\mathfrak{S}%
_{\mathbf{Z}}$. In Sect. \ref{Marker_NonEq-S}, we introduce the statistical
entropy formulation denoted by $\mathcal{S}$, and show its equivalence with
thermodynamic nonequilibrium entropy $S$ when the latter is a state function.
In Sect. \ref{Sec-ExtendedS-M-nieq}, we discuss the extended state space
needed for a state not in IEQ and show how it can be converted to
$\mathcal{M}_{\text{ieq}}$ by adding more internal variables to enlarge the
state space to $\mathfrak{S}_{\mathbf{Z}^{\prime}}\supset\mathfrak{S}%
_{\mathbf{Z}}$. In Sect. \ref{Sec-EntropyCalculation}, we give a simple
calculation and a brief introduction to chemical reaction model to change
$p_{k}$'s. In Sect. \ref{Sec-Applications}, we give many applications of using
$\mathfrak{S}_{\mathbf{Z}}$ and calculate the irreversible entropy generation
in $\mathcal{M}_{\text{ieq}}$. In Sect. V, we consider a $1$-d lattice
model\ appropriate for Tonks gas in continuum so that the statistical lattice
entropy can be calculated rigorously. We take the continuum limit and compare
the resulting entropy with the continuum entropy of the Tonks gas and obtain
an interesting result. In Sect. \ref{Marker_Jaynes}, we revisit Jaynes and
obtain his bound. \ A brief summary and discussion is presented in the final section.

\section{The Second Law and A Nonequilibrium State}

\subsection{Second Law}

Following deGroot and Prigogine, we write%
\begin{equation}
dS=d_{\text{e}}S+d_{\text{i}}S \label{Entropy-Partition}%
\end{equation}
during any infinitesimal physical process in $\Sigma_{\text{b}}$, where
$d_{\text{e}}S$ is the entropy exchange with the medium and $d_{\text{i}}S$ is
the irreversible entropy generated within $\Sigma_{\text{b}}$; see Fig.
\ref{Fig.Sys}. The second law states that $d_{\text{i}}S$ satisfies the
inequality%
\begin{equation}
d_{\text{i}}S\geq0; \label{Second_Law0}%
\end{equation}
the equality occurs for a reversible process. For an isolated system,
$d_{\text{e}}S\equiv0$. Hence, $dS_{0}=d_{\text{i}}S$ in any arbitrary process
and satisfies%
\begin{equation}
dS_{0}\geq0. \label{Second_Law}%
\end{equation}
We thus see that the second law statement in EQ. (\ref{Second_Law0}) is the
most general one and applies to any body during any physical process. The law
refers to the \emph{thermodynamic entropy}.

As the thermodynamic entropy is not measurable except when the process is
reversible, the second law remains useless as a computational tool. In
particular, it says nothing about the rate at which the irreversible entropy
increases. Therefore, it is useful to obtain a computational formulation of
the NEQ entropy. Several examples in Sect. \ref{Sec-Applications} show how
irreversible entropy generation can be measured by going to the enlarged state
space; see, for example, the discussion after Eq. (\ref{diS-dEin}). In
addition, the statistical entropy also proves useful to obtain a computational
scheme to determine the entropy by using the microstate probabilities derived
in Eq. (\ref{microstate probability}).

\subsection{Concept of a Nonequilibrium
State\label{Marker_Internal Equilibrium0}}

For a body in equilibrium, the entropy can be expressed as a function of its
observables (variables that can be controlled by an observer), as is easily
seen form the Gibbs fundamental relation in Eq. (\ref{Gibbs_Fundamental}). The
thermodynamic state $\mathcal{M}_{\text{eq}}$ in EQ remains the same unless it
is disturbed. Therefore, we can uniquely identify $\mathcal{M}_{\text{eq}}$ by
its observable $\mathbf{X}$. Accordingly, its equilibrium entropy
$S(\mathbf{X})$ is also expressed as a function of $\mathbf{X}$,
\textit{i.e.}, $S(\mathbf{X})$ is a \emph{state function}, and $\mathbf{X}%
$\ is the set of \emph{state variables}. We denote the state space associated
with $\mathbf{X}$ by $\mathfrak{S}_{\mathbf{X}}$.

The above conclusion is most certainly not valid for a body out of
equilibrium. Let us consider an isolated body $\Sigma_{\text{b}}=\Sigma_{0}$,
and use the suffix $0$ for the moment for clarity. If it is not in
equilibrium, its state $\mathcal{M}_{0}(t)$ will continuously change, which is
reflected in its entropy increase in time; this requires expressing $S$ as
$S(\mathbf{X}_{0},t)$ with an \emph{explicit time-dependence}, since
$\mathbf{X}_{0}=constant$ for $\Sigma_{0}$. The change in the entropy and the
state must come from the variations of additional variables, distinct from the
observables, that keep changing with time until $\Sigma_{0}$ comes to EQ as
explained elsewhere \cite{Gujrati-I,Gujrati-II,Gujrati-III}. These are known
as the \emph{internal variables }(sometimes also called hidden variables); see
Sect. \ref{Sec-InternalVariables} for how to identify them in some simple
situations. We should emphasize that the concept of internal variables and
their usefulness in NEQT has a long history. We refer the reader to an
excellent exposition of this topic in the monograph by Maugin \cite[see Ch.
4]{Maugin}. These variables cannot be controlled by the observer. Once the
body has come to equilibrium, the entropy has no explicit time-dependence and
becomes a state function. In this state, the entropy has its maximum possible
value for given $\mathbf{X}_{0}$.

\subsection{Internal Equilibrium States\label{Marker_Internal Equilibrium}}

We turn to a body again. We assume that there is a set $\boldsymbol{\xi}$
of\ additional variables, known as the internal variables\emph{ }(sometimes
also called hidden variables); see Sect. \ref{Sec-InternalVariables} for how
to identify them in some simple situations. We will refer to the variables in
$\mathbf{X}$ and $\boldsymbol{\xi}$\ as (\emph{nonequilibrium})\ \emph{state
variables} (see below for justification) and denote them collectively as
$\mathbf{Z}$ in the following, with $\mathfrak{S}_{\mathbf{Z}}$ for the
\emph{enlarged state space}. From Theorem 4 presented elsewhere
\cite{Gujrati-II}, it follows that with a proper choice of the number of
internal variables, the entropy can be written as $S(\mathbf{Z}(t))$ as a
\emph{state function} with no explicit $t$-dependence so it becomes unique.
The situation is now almost identical to that of a body in equilibrium:\ The
entropy is a function of
\[
\mathbf{Z}(t)=(E(t),\mathbf{W}(t))=(E(t),\mathbf{w}(t),\boldsymbol{\xi}(t))
\]
with no explicit time-dependence; here, we have introduced $\mathbf{W}(t)$ and
$\mathbf{w}(t)$ obtained by taking out $E(t)$ from $\mathbf{Z}(t)$ and
$\mathbf{X}(t)$, respectively, so that $\mathbf{W}(t)=(\mathbf{w}%
(t),\boldsymbol{\xi}(t))$.\ (We will see below that $\mathbf{W}$ determines
the work done by the body.) As $S$ is a state function, we can identify
$\mathbf{Z}(t)$\ as the set of NEQ state variables. Thus, $\mathcal{M}(t)$ can
also be uniquely specified by $\mathbf{Z}(t)$. This allows us to extend Eq.
(\ref{Gibbs_Fundamental}) to%
\begin{equation}
dS=\left(  \partial S/\partial\mathbf{Z}\right)  \cdot d\mathbf{Z}
\label{Gibbs_Fundamental_Extended}%
\end{equation}
in which the partial derivatives are related to the fields of the body:%
\begin{subequations}
\begin{equation}
\beta=1/T=\partial S/\partial E,\beta\mathbf{F}_{\text{w}}=-\partial
S/\partial\mathbf{W.} \label{Fields_Isolated0}%
\end{equation}
These fields will change in time unless the body has reached equilibrium. In
conventional notation,
\begin{equation}
\left(  \partial S/\partial V\right)  =\beta P,\cdots,\partial S/\partial
\boldsymbol{\xi}=-\beta\mathbf{A,} \label{Fields_Isolated}%
\end{equation}
where $P$ is the pressure and $\mathbf{A}$ the affinity corresponding to
$\boldsymbol{\xi}$; the missing terms $\cdots$ represent terms from the rest
of $\mathbf{X}$ besides $E$ and $V$. In EQ, $\mathbf{F}_{\text{w}}$ takes the
EQ value $\mathbf{F}_{\text{w}0}$ associated with $\widetilde{\Sigma}$.

We can invert Eq. (\ref{Gibbs_Fundamental_Extended}) to express $E$ in terms
of $S$ and $\mathbf{W}$:%
\end{subequations}
\begin{equation}
dE=TdS-\mathbf{F}_{\text{w}}\cdot d\mathbf{W,} \label{FirstLaw-Generalized}%
\end{equation}
where $\mathbf{F}_{\text{w}}=-\partial E/\partial\mathbf{W}$ is identified as
a force $\mathbf{F}_{\text{w}}$, and $\mathbf{W}$ as a "work variable." We
will call $\mathbf{F}_{\text{w}}$ the \emph{generalized force} as not all
components of $d\mathbf{W}$ represent displacement in space. This allows us to
identify the \emph{generalized work}
\begin{equation}
dW=\mathbf{F}_{\text{w}}\cdot d\mathbf{W=f}_{\text{w}}\cdot d\mathbf{w+A}\cdot
d\boldsymbol{\xi} \label{Generalized work}%
\end{equation}
as the work done \emph{by} the body; here $\mathbf{f}_{\text{w}}=-\partial
E/\partial\mathbf{w}$. As $\mathbf{f}_{\text{w}}\cdot d\mathbf{w}=PdV+\cdots$
in terms of $P$, $\cdots$ of the body, it must not be confused with the work
$d\widetilde{W}=P_{0}d\widetilde{V}+\cdots$ done \emph{on} $\Sigma_{\text{b}%
}=\Sigma$ by the medium $\widetilde{\Sigma}$. Note that the EQ value
$\mathbf{A}_{0}=0$ of the affinity represents the affinity of $\widetilde
{\Sigma}$ so $\mathbf{A}_{0}\cdot d\boldsymbol{\xi}\equiv0$ contributes
nothing to the work done on $\Sigma$ . We use $d_{\text{e}}W=-d\widetilde{W}$
to denote this \emph{external} or exchange work done by $\Sigma$\ against
$\widetilde{\Sigma}$. Their difference is the irreversible work $d_{\text{i}%
}W$ \cite{Woods,Gujrati-II,Kestin} in $\Sigma$\ so that
\begin{equation}
dW=d_{\text{e}}W+d_{\text{i}}W. \label{Work-Partition}%
\end{equation}
This partition is similar to the one in Eq. (\ref{Entropy-Partition}) for the
entropy, and has a similar explanation: $d_{\text{e}}W$ is the exchange work
done by $\Sigma$ on $\widetilde{\Sigma}$, and $d_{\text{i}}W$ is the
irreversible or internal work generated within $\Sigma$. In contrast, $dW$ is
the net work done by $\Sigma$. We wish to remind the reader that the concepts
of two different kinds of NEQ work in classical thermodynamics have been well
known but not with the present interpretation; see for example, \cite[Sec.
3.3]{Woods}. Thus, care must be exercised in distinguishing the two works in a
NEQ process, which is not always the case \cite{Gujrati-MistakenIdentity}. The
above partition first proposed in \cite{Gujrati-II} formalizes this
distinction as a natural extension of Eq. (\ref{Entropy-Partition}). Indeed,
the above partitioning can be done for any extensive quantity $Z$ pertaining
to the body as shown in Fig. \ref{Fig.Sys}:%
\begin{equation}
dZ=d_{\text{e}}Z+d_{\text{i}}Z, \label{Zee-Partition}%
\end{equation}
having the conventional interpretation: $d_{\text{e}}Z$ is the part exchanged
with the medium and $d_{\text{i}}Z$ is the irreversible part generated within
the system. Their sum makes up the net change $dZ$ in the quantity. It is well
known that $d_{\text{i}}E=0$ and $d_{\text{i}}V=0$ for the simple reason that
no internal process can change $E$ and $V$. However, $d_{\text{i}}N\neq0$ when
there is chemical reaction within the system.

The expression for $d_{\text{e}}W$ in the general case is obtained by
replacing $\mathbf{f}_{\text{w}}$ by its EQ value $\mathbf{f}_{\text{w}0}$ and
$\mathbf{A}$ by its EQ value $\mathbf{A}_{0}=0$ associated with $\widetilde
{\Sigma}$, and replacing $d\mathbf{w=}d_{\text{e}}\mathbf{w+}d_{\text{i}%
}\mathbf{w}$ by its exchange contribution $d_{\text{e}}\mathbf{w}$. Thus,%
\begin{equation}
d_{\text{e}}W=\mathbf{f}_{\text{w}0}\cdot d_{\text{e}}\mathbf{w}%
=P_{0}dV+\cdots\label{ExchangeWork}%
\end{equation}
as noted above. This finally determines the irreversible work%
\begin{equation}
d_{\text{i}}W=(\mathbf{f}_{\text{w}}-\mathbf{f}_{\text{w}0})\cdot d_{\text{e}%
}\mathbf{w+f}_{\text{w}}\cdot d_{\text{i}}\mathbf{w+A}\cdot d\boldsymbol{\xi
}\geq0\boldsymbol{,} \label{IrreversibleWork}%
\end{equation}
which also establishes the fact that internal variables only contribute to
internal processes; they are never involved in any exchange processes. The
inequality is a consequence of the second law as is easily proven
\cite{Gujrati-Hierarchy}; we refer the reader to this for further details.

The exchange heat $d_{\text{e}}Q=T_{0}d_{\text{e}}S$ is used in the
traditional way of writing the first law%
\begin{subequations}
\begin{equation}
dE=d_{\text{e}}Q-d_{\text{e}}W \label{FirstLaw-MI}%
\end{equation}
in terms of exchange quantities with $\widetilde{\Sigma}$. Using the
definition of $dW$, we have from Eq. (\ref{FirstLaw-Generalized})%
\begin{equation}
dE=dQ-dW, \label{FirstLaw-SI}%
\end{equation}
where we have introduced the \emph{generalized heat}
\end{subequations}
\begin{equation}
dQ=TdS, \label{dQ-dS}%
\end{equation}
which is partitioned similar to $dW$ above as
\begin{equation}
dQ=d_{\text{e}}Q+d_{\text{i}}Q. \label{Heat-Partition}%
\end{equation}
Again, the interpretation of $d_{\text{i}}Q$ is similar:\ it is the
irreversible or internal heat generated within $\Sigma$. Thus, $dQ$ is the net
heat. Comparing the two formulations of $dE$ above, we conclude that%
\begin{equation}
d_{\text{i}}Q=d_{\text{i}}W. \label{diQ-diW}%
\end{equation}

It should be clear from the above that the use of quantities pertaining to the
body alone, which we call \emph{system-intrinsic }(SI-)\emph{ }quantities,
captures the entire irreversibility in the NEQ body. As an example, Eq.
(\ref{FirstLaw-SI}), which originated from the Gibbs fundamental relation, is
in terms of SI-quantities $dQ$ and $dW$. Thus, it also represents a new
version of the first law but in terms of the SI-quantities and fully captures
the irreversibility in the body. In contrast, the exchange quantities are
determined by the medium, which we call \emph{medium-intrinsic }(MI-)\emph{
}quantities, that are oblivious to what is going on within the system. Thus,
the first law in Eq. (\ref{FirstLaw-MI}), although applicable to any process,
cannot provide any information of the irreversibility in the body. This shows
the usefulness of Eq. (\ref{FirstLaw-SI}), and the use of SI-quantities.

We now have a complete NEQ thermodynamics in terms of SI-quantities, which
contain all the irreversible components as seen above. We have identified this
thermodynamics by MNEQT, with M standing for the use of MI-quantities
\cite{Gujrati-LangevinEq}. To verify that we have captured entire
irreversibility, we determine $d_{\text{i}}Q=dQ-d_{\text{e}}Q$, which turns
out to be%
\begin{subequations}
\begin{equation}
d_{\text{i}}Q=(T-T_{0})d_{\text{e}}S+Td_{\text{i}}S. \label{diQ}%
\end{equation}
For the simple case of $\mathbf{W}=(V,\xi)$, we have $dW=PdV+Ad\xi$. The
exchange work is $d_{\text{e}}W=P_{0}dV$ so%
\begin{equation}
d_{\text{i}}W=(P-P_{0})dV+Ad\xi, \label{diW}%
\end{equation}
from which we also obtain by using Eq. (\ref{diQ-diW}) the following
expression for the irreversible entropy%
\end{subequations}
\begin{equation}
Td_{\text{i}}S=(T_{0}-T)d_{\text{e}}S+(P-P_{0})dV+Ad\xi
\label{Irreversible EntropyGeneration-Complete}%
\end{equation}
for this simple case. The first two contributions are due to exchanges with
$\widetilde{\Sigma}$, and the last term is from the internal variable; it
should be replaced by $\mathbf{A}\cdot d\boldsymbol{\xi}$ in the general case.
As we see from Eq. (\ref{IrreversibleWork}), there is an additional
contribution, not seen above, from $\mathbf{f}_{\text{w}}\cdot d_{\text{i}%
}\mathbf{w}$. Each contribution in $d_{\text{i}}S$ must be nonnegative in
accordance with the second law.

As $\mathbf{Z}(t)$\ changes in time, $\mathcal{M}(t)$ changes, but at each
instant the NEQ entropy as a state function, has a maximum possible value for
given $\mathbf{Z}(t)$ even though $\mathcal{M}(t)\neq\mathcal{M}_{\text{eq}}$.
We have identified this particular state as an \emph{internal equilibrium
state }(IEQ) \cite{Gujrati-I,Gujrati-II,Gujrati-III} and express it as
$\mathcal{M}_{\text{ieq}}$. In $\mathfrak{S}_{\mathbf{Z}}$, $\mathcal{M}%
_{\text{ieq}}$ is uniquely described. There are many states that are not in
IEQ in that they are not unique in $\mathfrak{S}_{\mathbf{Z}}$. We will denote
such states by $\mathcal{M}_{\text{nieq}}$ (\emph{nonIEQ} state), and use an
arbitrary state $\mathcal{M}_{\text{arb}}$ to denote any state by not
specifying the state space $\mathfrak{S}$.

We clarify this point. If we do not use $\boldsymbol{\xi}$ for a NEQ
$\mathcal{M}$, it is not unique in $\mathfrak{S}_{\mathbf{X}}$. Then its
entropy cannot be a state function in $\mathfrak{S}_{\mathbf{X}}$, and must be
expressed as $S(\mathbf{X},t)$. This explains the importance of
$\boldsymbol{\xi}$; it allows us to deal with a state function entropy
$S(\mathbf{Z})$ by choosing an \emph{appropriate} number of internal
variables. Throughout this chapter, we will mostly deal with IEQ states. But
our discussion will cover all states ($\mathcal{M}_{\text{arb}}$) in any
$\mathfrak{S}$.

As a state function, $S(\mathbf{Z})$ shares many of the properties of EQ
entropy $S(\mathbf{X})$ \cite{Gujrati-II}:

\begin{enumerate}
\item \textit{Maximum}: $S(\mathbf{Z})$ is the maximum possible value of the
NEQ entropy in $\mathfrak{S}_{\mathbf{Z}}$ for a given $\mathbf{Z}$.

\item \textit{No memory} -Its value also does not depend on how the system
arrives in $\mathcal{M}_{\text{ieq}}\equiv\mathcal{M}(\mathbf{Z})$,
\textit{i.e., }whether\textit{ }it arrives there from another IEQ state
$\mathcal{M}_{\text{ieq}}$ or a non-IEQ state $\mathcal{M}_{\text{nieq}}$.
Thus, it has no memory of the earlier state.
\end{enumerate}

There are states ($\mathcal{M}_{\text{nieq}}$) in $\mathfrak{S}_{\mathbf{Z}}$
for which the entropy is not a state function. They possess memory of the
initial states in $\mathfrak{S}_{\mathbf{Z}}$ with the entropy $S(\mathbf{Z}%
,t)$ retaining an explicit time-dependence. In this case, we need to enlarge
the state space to $\mathfrak{S}_{\mathbf{Z}^{\prime}}$ by including
additional internal variables as shown elsewhere \cite{Gujrati-II} and latter
in Sect. \ref{Sec-M_nieq}. Then $\mathcal{M}_{\text{nieq}}(\mathbf{Z})$ in
$\mathfrak{S}_{\mathbf{Z}}$\ turns into $\mathcal{M}_{\text{ieq}}%
(\mathbf{Z}^{\prime})$ in $\mathfrak{S}_{\mathbf{Z}^{\prime}}$. In
$\mathfrak{S}_{\mathbf{Z}^{\prime}}$, the derivatives in Eq.
(\ref{Fields_Isolated}) with $\mathbf{Z}$ replaced by $\mathbf{Z}^{\prime}%
$\ can again be identified as field variables like, temperature, pressure,
etc. We will explain in Sect. \ref{Sec-Choice-S_Z} how to determine the
correct NEQ state space based on the way experiments are performed.

It may appear to a reader that the concept of entropy being a state function
is very restrictive. This is not the case as this concept, although not
recognized by several workers, is implicit in the literature where the
relationship of the thermodynamic entropy with state variables is
investigated. To appreciate this, we observe that the entropy of a body in
internal equilibrium\cite{Gujrati-I,Gujrati-II} is given by the Boltzmann
formula%
\begin{equation}
S(\mathbf{Z}(t))=\ln W(\mathbf{Z}(t)), \label{Boltzmann_S_Extended}%
\end{equation}
in terms of the number of microstates corresponding to $\mathbf{Z}(t)$. In
classical nonequilibrium thermodynamics \cite{deGroot}, the entropy is always
taken to be a state function. In the Edwards approach \cite{Edwards} for
granular materials, all microstates are equally probable as is required for
the above Boltzmann formula. Bouchbinder and Langer \cite{Langer} assume that
the nonequilibrium entropy is given by Eq. (\ref{Boltzmann_S_Extended}).
Lebowitz \cite{Lebowitz} also takes the above formulation for his definition
of the nonequilibrium entropy. As a matter of fact, we are not aware of any
work dealing with entropy computation that does not assume the nonequilibrium
entropy to be \ a state function. This does not, of course, mean that all
states of a system are internal equilibrium states. For states that are not in
internal equilibrium, the entropy is not a state function so that it will have
an explicit time dependence. But this can be avoided by further enlarging the
space of internal variables that results in $\mathbf{Z}^{\prime}$\ discussed
above. The choice of how many internal variables are needed will depend on
experimental time scales and has been answered in generality earlier in
\cite{Gujrati-Hierarchy}, and is briefly summarized in Sect.
\ref{Sec-ExtendedS-M-nieq}.

\section{Internal Variables\label{Sec-InternalVariables}}

It should again be stated that in order to capture a NEQ process, internal
variables are usually \emph{necessary}; see Remark \ref{Remark-IsolatedSystem}%
. While a point in $\mathfrak{S}_{\mathbf{X}}$ represents $\mathcal{M}%
_{\text{eq}}$, we need to use the enlarged state space $\mathfrak{S}%
_{\mathbf{Z}}$ in which a point represents $\mathcal{M}$. As internal
variables are not required in EQ, they must no longer be independent of the
observables in $\mathfrak{S}_{\mathbf{X}}$. Consequently, their affinities
(see Eq. (\ref{Fields_Isolated}) for $\mathbf{A}$) vanish in EQ. It is common
to define the internal variables so their EQ values vanish, but it is not
necessary. We now discuss various scenarios where they are required for a
proper NEQ thermodynamic consideration.

\subsection{A Two-level System}

Consider a NEQ body $\Sigma_{\text{b}}=\Sigma_{0}$ of $N$ particles such as
Ising spins, each of which can be in two levels, forming an isolated system
$\Sigma_{0}$ of volume $V$. Let $\rho_{l}$ and $e_{l}(V),l=1,2$ denote the
probabilities and energies of the two levels of a particle in a NEQ state so
that $\rho_{1},\rho_{2}$ keep changing. We have assumed that $e_{l}(V)$
depends on the observable $V$ only, which happens to be constant for
$\Sigma_{\text{b}}=\Sigma_{0}$. We have $e=\rho_{1}e_{1}(V)+\rho_{2}e_{2}(V)$
for the average energy per particle, which is also a constant. We have
\[
d\rho_{1}+d\rho_{2}=0
\]
as a consequence of $\rho_{1}+\rho_{2}=1$. Using $de=0$, we also get
\[
d\rho_{1}+d\rho_{2}e_{2}/e_{1}=0,
\]
which, for $e_{1}\neq e_{2}$, is inconsistent with the first equation (unless
$d\rho_{1}=0=d\rho_{2}$, which corresponds to EQ). Thus, $e_{l}(V)$ cannot be
treated as constant in evaluating $de$. In other words, there must be an extra
dependence in $e_{l}$ so that%
\[
e_{1}d\rho_{1}+d\rho_{2}e_{2}+\rho_{1}de_{1}+\rho_{2}de_{2}=0,
\]
and the inconsistency is removed. This extra dependence must be due to
\emph{independent} internal variables that are not controlled from the outside
so they continue to relax in $\Sigma_{\text{b}}$ as it approaches EQ. Let us
imagine that there is a single internal variable $\xi$ so that we can express
$e_{l}$ as $e_{l}(V,\xi)$ in which $\xi$ continues to change as the system
comes to equilibrium. The above equation then relates $d\rho_{1}$ and $d\xi$;
they both vanish simultaneously as EQ is reached. We also see that without any
$\xi$, the isolated system cannot equilibrate in accordance with Remark
\ref{Remark-IsolatedSystem}.

\subsection{A Many-level System}

The above discussion is easily extended to a $\Sigma$ with many energy levels
of a particle with the same conclusion that at least a single internal
variable is required to express $e_{l}=e_{l}(V,\xi)$ for each level $l$. We
can also visualize the above system in terms of microstates. A microstate
$\mathfrak{m}_{k}$ refers to a particular distribution of the $N$ particles in
any of the levels with energy $E_{k}=%
{\textstyle\sum\nolimits_{l}}
N_{l}e_{l}$, where $N_{l}$ is the number of particles in the $l$th level, and
is obviously a function of $N,V,\xi$ so we will express it as $E_{k}(N,V,\xi
)$. This makes the average energy of the system also a function of $N,V,\xi$,
which we express as $E(N,V,\xi)$.

\subsection{Disparate Degrees of freedom}

In classical statistical mechanics, the kinetic and potential energies $K$ and
$U$, respectively, are functions of independent variables. Only their sum
$K+U=E$ can be controlled from the outside, but not individually. Thus, one of
them can be treated as an internal variable. In a NEQ states, each term can
have its own temperature. Only in EQ, do they have the same temperature.

This has an important consequence for glasses \cite{Gutzow,Nemilov}, where the
translational degrees of freedom come to EQ with the heat bath at $T_{0}$
faster than the configurational degrees of freedom, which have a different
temperature than $T_{0}$. The disparity cannot be controlled by the observer
so it plays the role of an internal variable.

Consider a collection of semiflexible polymers in a solution on a lattice. The
interaction energy $E$ consists of several additive terms as discussed in
\cite[Eq. (40)]{Gujrati-II}: the interaction energy $E_{\text{ps}}$\ between
the polymer and the solvent, the interaction energy $E_{\text{ss}}$\ between
the solvent, the interaction energy $E_{\text{pp}}$\ between polymers. Only
the total $E$ can be controlled from the outside so the remaining terms
determine several internal variables.%

\begin{figure}
[ptb]
\begin{center}
\includegraphics[
height=1.0914in,
width=2.5071in
]%
{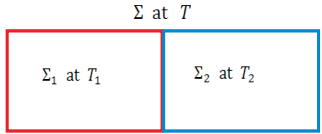}%
\caption{A composite system $\Sigma$ consisting of two identical subsystems
$\Sigma_{1}$ at temperature $T_{1}$ and $\Sigma_{2}$ at temperature $T_{2}$.
It will be seen later in Sec. \ref{Sec-Composite System} that the
thermodynamic temperature of $\Sigma$ can be defined as $T$ given by Eq.
(\ref{beta-A-Composite}). The irreversibility in $\Sigma$ requires one
internal variable $\xi$ given in Eq. (\ref{Internal Variable-1}).}%
\label{Fig-CompositeSystem}%
\end{center}
\end{figure}

\subsection{Nonuniformity \label{Sec-Nonuniformity}}

In the examples above, the internal variables are not due to any spatial
inhomogeneity. An EQ system is uniform. Thus, the presence of $\xi$ suggests
some sort of nonuniformity in the system. To appreciate its physics, we
consider a slightly different situation below as possible examples of
nonuniformity. We will consider a system $\Sigma$ for simplicity.

(a) We consider as a simple NEQ example a composite isolated system $\Sigma$,
see Fig. \ref{Fig-CompositeSystem}, consisting of two subsystems $\Sigma_{1}$
and $\Sigma_{2}$ of identical volumes and numbers of particles but at
different temperatures $T_{1}$ and $T_{2}$ at any time $t<\tau_{\text{eq}}$
before EQ is reached at $t=\tau_{\text{eq}}$ so the subsystems have different
time-dependent energies $E_{1}$ and $E_{2}$, respectively. We assume a
diathermal wall separating $\Sigma_{1}$ and $\Sigma_{2}$. Treating each
subsystem in EQ at each $t$, we write their entropies as $S_{1}(E_{1}%
,V/2,N/2)$ and $S_{2}(E_{2},V/2,N/2)$, which we simply show as $S_{1}(E_{1})$
and $S_{2}(E_{2})$ as we will not let their volumes and particles numbers
change. The entropy $S$ of $\Sigma$ is a function of $E_{1}$ and $E_{2}$.
Obviously, $\Sigma$ is in a NEQ state at each $t<\tau_{\text{eq}}$. From
$E_{1}$ and $E_{2}$, we form two independent combinations%

\begin{equation}
E=E_{1}+E_{2},\xi=E_{1}-E_{2}, \label{Internal Variable-1}%
\end{equation}
so that we can express the entropy as $S(E,\xi)$ for $\Sigma$ treated as a
black box $\Sigma_{\text{B}}$; we do not need to know about its interior (its
inhomogeneity) anymore. Here, $\xi$ plays the role of an internal variable,
which continues to relax\ towards zero as $\Sigma$ approaches EQ. For given
$E$ and $\xi$,\ $S(E,\xi)$ has the \ maximum possible values since both
$S_{1}$ and $S_{2}$ have their maximum value. As we will see below, this is
the idea behind the concept of \emph{internal equilibrium} in which $S(E,\xi)$
is a state function of state variables and continues to increase as $\xi$
decreases and vanishes in EQ. In this state, $S(E,\xi=0)$ has the maximum
possible value for fixed $E$ so it becomes a state function. This case and its
various extensions are investigated in MNEQT in Sect.
\ref{Sec-Composite System}.

(b) We can easily extend the model to include four identical subsystems of
fixed and identical volumes and numbers of particles, but of different
energies $E_{1},E_{2},E_{3}$, and $E_{4}$. Instead of using these $4$
independent variables, we can use the following four independent combinations%
\begin{align}
E  &  =E_{1}+E_{2}+E_{3}+E_{4}=\text{constant},\nonumber\\
\xi &  =E_{1}+E_{2}-E_{3}-E_{4},\nonumber\\
\text{ }\xi^{\prime}  &  =E_{1}-E_{2}+E_{3}-E_{4},\nonumber\\
\xi^{\prime\prime}  &  =E_{1}-E_{2}-E_{3}+E_{4}, \label{Internal Variable-2}%
\end{align}
to express the entropy of $\Sigma$ as $S(E,\xi,\xi^{\prime},\xi^{\prime\prime
})$. The pattern of extension for this simple case of energy inhomogeneity. is evident.

(c) We make the model a bit more interesting by allowing the volumes $V_{1}$
and $V_{2}$ to also vary as $\Sigma$ equilibrates. Apart from the internal
variable $\xi$, we require another internal variable $\xi^{\prime}$ to form
two independent combinations
\begin{equation}
V=V_{1}+V_{2}=\text{constant},\text{ }\xi^{\prime}=V_{1}-V_{2}
\label{Internal Variable-3}%
\end{equation}
so that we can use $S(E,V,\xi,\xi^{\prime})\doteq S_{1\text{eq}}(E_{1}%
,V_{1})+S_{2\text{eq}}(E_{2},V_{2})$ for the entropy of $\Sigma$ in terms of
the entropies of $\Sigma_{1}$ and $\Sigma_{2}$.

(d) In the above examples, we have assumed the subsystems to be in EQ. We now
consider when the subsystems are in IEQ. We consider the simple case of two
subsystems $\Sigma_{1}$ and $\Sigma_{2}$ of identical volumes and numbers of
particles. Each subsystem is in different IEQ states described by $E_{1}%
,\xi_{1}$ and $E_{2},\xi_{2}$. We now construct four independent combinations%
\begin{align}
E  &  =E_{1}+E_{2}=\text{constant},\xi=E_{1}-E_{2},\nonumber\\
\xi^{\prime}  &  =\xi_{1}+\xi_{2},\xi^{\prime\prime}=\xi_{1}-\xi_{2},
\label{Internal Variable-4}%
\end{align}
which can be used to express the entropy of $\Sigma$ as $S(E,\xi,\xi^{\prime
},\xi^{\prime\prime})$.

(e) The example in (a) can be easily extended to the case of expansion and
contraction by replacing $E,E_{1}$, and $E_{2}$ by $N,N_{\text{L}}$, and
$N_{\text{R}}$ to describe the diffusion of particles from the left (L) region
to the right (R) region containing $N_{\text{L}}$ and $N_{\text{R}}$
particles, respectively. The role of $\beta$ and $E$, etc. are played by
$\beta\mu$ and $N$, etc.

It should be clear from the above examples that the choice of the number
$n^{\ast}$ of internal variables is determined uniquely by a particular
modeling of $\Sigma_{\text{b}}$. The number is fixed by matching the number of
independent variables needed for the subsystems to be exactly the number of
variables needed for $\Sigma_{\text{b}}$. This uniqueness ensures that the
entropy of $\Sigma_{\text{b}}$ is uniquely determined by the sum of the
entropies of the subsystems. Ensuring subsystems to be in EQ so that their
entropies are known, the uniqueness implies that the entropy of $\Sigma
_{\text{b}}$ is also known as a function of all the variables of subsystems.

\section{General Formulation of the Statistical Entropy\label{Marker_NonEq-S}%
\ \ \ }

In this section, we will use $\mathcal{S}$ for the statistical entropy to
distinguish it from the thermodynamic entropy $S$. We provide a very general
formulation of $\mathcal{S}$ for a general body $\Sigma_{\text{b}}$, which
will be shown to be identical to the thermodynamic entropy $S$ by appealing to
the third law. As a consequence, this will also demonstrate that the entropy
in general is a \emph{statistical average}. We consider a state $\mathcal{M}%
(t)\equiv\mathcal{M}(\mathbf{Z}(t))$ of $\Sigma_{\text{b}}$\ at a given
instant $t$. In the following, we suppress $t$ unless necessary. The state
$\mathcal{M}(\mathbf{Z})$ refers to the sets of microstates $\mathbf{m}%
=\left\{  \mathfrak{m}_{k}\right\}  $\ and their probabilities $\mathbf{p}%
=\left\{  p_{k}\right\}  $. The microstates are determined by the Hamiltonian
of the body, whose value will determine their energies $E_{k}$ for a given set
$\mathbf{W}(t)$. Thus, $\mathfrak{m}_{k}$ is specified by $(E_{k}%
(t),\mathbf{W}(t))$, and may not uniquely specify $\mathcal{M}(t)$ unless we
are in the appropriate state space where they become uniquely specified. In
the following, we will not require $\mathcal{M}(t)$ to be unique in the state
space $\mathfrak{S}_{\mathbf{Z}}$. We will denote $\mathbf{Z}(t)$ by
$\overline{\mathbf{Z}}$ so that we can separate out the explicit variation due
to $t$, and simply use $\mathcal{M}$ in the following.

For the computation of combinatorics, the probabilities are handled in the
following abstract way. We consider a large number $\mathcal{N=C}%
W(\overline{\mathbf{Z}})$ of independent \emph{replicas} or \emph{samples} of
$\Sigma_{\text{b}}$, with $\mathcal{C}$\ some large constant integer and
$W(\overline{\mathbf{Z}})$\ the number of distinct microstates $\mathfrak{m}%
_{k}$. The samples should be thought of as identically prepared experimental
samples \cite{Gujrati-Symmetry}.

\subsection{Simply Connected Sample
Space\ \ \ \ \ \ \ \ \ \ \ \ \ \ \ \ \ \ \ \ \ \ \ \ \ \ \ \ \ \ \ \ \ \ \ \ \ \ \ \ \ \ \ \ \ \ \ \ \ \ \ \ \ \ \ \ \ \ \ \ \ \ \ \ \ \ \ \ \ \ \ \ \ \ \ \ \ \ \ \ \ \ \ \ \ \ \ \ \ \ \ \ \ \ \ \ \ \ \ \ \ \ \ \ \ \ \ \ \ \ \ \ \ \ \ \ \ \ \ \ \ \ \ \ \ \ \ \ \ \ \ \ \ \ \ \ \ \ \ \ \ \ \ \ \ \ \ \ \ \ \ \ \ \ \ \ \ \ \ \ }%

\subsubsection{An Isolated Body\label{Marker_Probabilities}}

We assume that $\Gamma(\overline{\mathbf{Z}})\supset$ $\Gamma(\overline
{\mathbf{X}})$ is simply connected in this section.\ Let $\mathcal{N}_{k}(t)$
denote the number of $k$th samples (samples in the $\mathfrak{m}_{k}%
$-microstate) so that%
\begin{equation}
0\leq p_{k}(t)=\mathcal{N}_{k}(t)/\mathcal{N}\leq1;\ \
{\textstyle\sum\limits_{k=1}^{W(\overline{\mathbf{Z}})}}
\mathcal{N}_{k}(t)=\mathcal{N}. \label{sample_probability}%
\end{equation}
The above sample space is a generalization of the \emph{ensemble} introduced
by Gibbs, except that the latter is restricted to an equilibrium body in
$\mathfrak{S}_{\mathbf{X}}$, whereas our sample space refers to the body in
$\mathfrak{S}_{\mathbf{Z}}$ in any arbitrary state for which $p_{k}$ may be
time-dependent. The \emph{ensemble average} of some quantity $\mathcal{Z}%
$\ over these samples is given by%
\begin{equation}
\left\langle \mathcal{Z}\right\rangle \equiv%
{\textstyle\sum\limits_{k=1}^{W(\overline{\mathbf{Z}})}}
p_{k}(t)\mathcal{Z}_{k},\ \
{\textstyle\sum\limits_{k=1}^{W(\overline{\mathbf{Z}})}}
p_{k}(t)\equiv1, \label{Ensemble_Average}%
\end{equation}
where $\mathcal{Z}_{k}$ is the value of $\mathcal{Z}$ in $\mathfrak{m}_{k}$.
This definition is identical to the Gibbs ensemble average in Eq.
(\ref{Ensemble Av}).

The samples are, by definition, \emph{independent} of each other so that there
are no correlations among them. Because of this, we can treat the samples to
be the outcomes of some random variable, the state $\mathcal{M}$. This
independence property of the outcomes is crucial in the following, and does
not imply that they are equiprobable. The number of ways $W$ to arrange the
$\mathcal{N}$ samples into $W(\overline{\mathbf{Z}})$ distinct microstates is%
\begin{equation}
\mathcal{W\equiv N}!/%
{\textstyle\prod\limits_{k}}
\mathcal{N}_{k}(t)!. \label{Combinations}%
\end{equation}
Taking its natural log to obtain an \emph{additive} quantity per sample, and
in accordance with Boltzmann's postulate in Eq. (\ref{Boltzmann_S}),%
\begin{equation}
\mathcal{S}\equiv\ln\mathcal{W}/\mathcal{N},
\label{Ensemble_entropy_Formulation}%
\end{equation}
and using Stirling's approximation, we see easily that the \emph{statistical
entropy} $\mathcal{S}$, which we hope to identify later with the entropy
$S(\overline{\mathbf{Z}})$ of $\Sigma_{\text{b}}$, can be written as the
average of the negative of%
\begin{equation}
\eta_{k}(t)\equiv\ln p_{k}(t), \label{Index-Prob}%
\end{equation}
what Gibbs \cite{Gibbs} calls the \emph{index of probability:}
\begin{equation}
\mathcal{S}(\overline{\mathbf{Z}},t)\equiv-\left\langle \eta(t)\right\rangle
\equiv-%
{\textstyle\sum\limits_{k=1}^{W(\overline{\mathbf{Z}})}}
p_{k}(t)\ln p_{k}(t), \label{Gibbs_Formulation}%
\end{equation}
where we have also shown an explicit time-dependence, which is distinct from
the implicit time-dependence in $\overline{\mathbf{Z}}$. The explicit
time-dependence in $\mathcal{S}(\overline{\mathbf{Z}},t)$ merely reflects the
fact that it is not a state function and that $\mathcal{M}$ is not uniquely
specified in $\mathfrak{S}_{\mathbf{Z}}$. The above derivation is based on
fundamental principles and the Boltzmann hypothesis and does not require the
body to be in equilibrium; therefore, it is always applicable. To the best of
our knowledge, even though such an expression has been extensively used in the
literature, it has been used \emph{without} any derivation; one simply appeals
to this form by appealing to the information entropy \cite{Shannon,Jaynes0};
however, see Sect. \ref{marker_Summary}.Thus, Eq. (\ref{Gibbs_Formulation}) is
a generalization of Eq. (\ref{Gibbs_S}) to the general case, and thus
justifies it for an arbitrary $\mathcal{M}$.

\begin{remark}
The statistical entropy $\mathcal{S}$ appears as an instantaneous ensemble
average, see Eq. (\ref{Ensemble Av}) or (\ref{Ensemble_Average}). This average
should be contrasted with a \emph{temporal average} in which a quantity
$\boldsymbol{\varphi}$ is considered as the average over a long period
$\tau_{0}$ of time%
\[
\boldsymbol{\varphi}=\frac{1}{\tau_{0}}%
{\textstyle\int\nolimits_{0}^{\tau_{0}}}
\boldsymbol{\varphi}(t)dt,
\]
where $\boldsymbol{\varphi}(t)$ is the value of $\boldsymbol{\varphi}$ at time
$t$ \cite{Landau}. For $\mathcal{M}_{\text{eq}}$, both definitions give the
same result provided ergodicity holds. The physics of this average is that
$\mathbf{\varphi}(t)$ at $t$ represents one of the microstate of
$\mathcal{M}_{\text{eq}}$. As $\mathcal{M}_{\text{eq}}$\ is invariant in time,
these microstates belong to $\mathcal{M}_{\text{eq}}$, and the time average is
the same as the ensemble average if ergodicity holds. However, for a NEQ state
$\mathcal{M}(t)$, which continuously changes with time, the temporal average
is not physically meaningful as the microstate probabilities at time $t$
corresponds to $\mathcal{M}(t)$ and not to $\mathcal{M}(t=0)$ in that the
probabilities and $\mathbf{Z}$ are different in the two states. Only the
ensemble average makes any sense at any time as was first pointed out in
\cite{Gujrati-Residual} because $p_{k}$'s correspond to $\mathcal{M}(t=0)$.
Because of this, we only consider ensemble averages in this chapter.
\end{remark}

Because of its similarity in form with $S_{\text{G}}^{(\text{c})}$ in Eq.
(\ref{Gibbs_S}), we will refer to $\mathcal{S}(\overline{\mathbf{Z}},t)$
simply as the Gibbs or the statistical entropy from now on. The distinction
between $\mathcal{S}$ and $S$ should be emphasized. The latter appears in the
Gibbs fundamental relation that relates the energy change $dE$ with the
entropy change $dS$ as is well known in classical thermodynamics, see Eqs.
(\ref{FirstLaw-Generalized}) and (\ref{FirstLaw-SI}) in $\mathfrak{S}%
_{\mathbf{Z}}$. The concept of microstates is irrelevant for this relation,
which is purely thermodynamic. On the other hand, $\mathcal{S}$ is solely
determined by $\left\{  \mathfrak{m}_{k}\right\}  $, and is a statistical
quantity which may not always satisfy Eq. (\ref{FirstLaw-Generalized}) in
$\mathfrak{S}_{\mathbf{Z}}$.

To identify $\mathcal{S}(\overline{\mathbf{Z}},t)$ with the NEQ thermodynamic
entropy $S$ requires the following additional steps:

\begin{enumerate}
\item[(1)] It is necessary to establish that $\mathcal{S}(\overline
{\mathbf{Z}},t)$ satisfies Eq. (\ref{Second_Law}).

\item[(2)] For body in canonical equilibrium, it is necessary to establish
that $\mathcal{S}(t)$ is identical to the equilibrium thermodynamic entropy
given by $S_{\text{G}}^{(\text{c})}$ \cite{Gibbs,Landau}.

\item[(3)] It is necessary to show that $\mathcal{S}(\overline{\mathbf{Z}})$
is identical to $S(\overline{\mathbf{Z}})$ for $\mathcal{M}_{\text{ieq}}$.

\item[(4)] It is necessary to show that $\mathcal{S}(\overline{\mathbf{Z}},t)$
is identical to $S(\overline{\mathbf{Z}},t)$ for $\mathcal{M}_{\text{nieq}}$.
\end{enumerate}

There are several proofs available in the literature
\cite{Tolman,Rice,Jaynes,Gujrati-Residual,Gujrati-Symmetry} for (1). Here, we
give a simple proof of it in Sect. \ref{Sec-SecondLawProof}. We will prove (2)
and (3) in Sect. \ref{Sec-MicrostateProbabilities}, where we follow closely
their justification given earlier \cite{Gujrati-Entropy1,Gujrati-Entropy2}. We
prove (4) in Sect. \ref{Sec-M_nieq}.

A word of caution must be offered. If $S$ is not a state function, it cannot
be measured or computed. Thus, while $\mathcal{S}$ can be computed in
principle in all cases, there is no way to compare its value with $S$ in all
cases. Thus, no comment can be made about their relationship in general. We
merely conjecture with respect to (4) that as the two entropies are the same
when the thermodynamic entropy is a state function, it is no different from
its statistical analog even when it is not a state function by following, in
principle, the procedure described in Sect. \ref{Sec-M_nieq}.

This allows us to identify $\mathcal{S}$ as the \emph{statistical entropy}
formulation of the thermodynamic entropy $S$ in all cases; see Proposition
\ref{Proposition-General-MNEQT}. Indeed, we use $\mathcal{S}$, which is
defined for any arbitrary state, to define the thermodynamic entropy $S$ in
all cases. From now on, we will not make any distinction between them. We
summarize this conclusion by the following

\begin{remark}
Because of this equivalence, we will no longer make any distinction between
the statistical entropy and the thermodynamic entropy and will use the
standard notation $S$\ for both of them, unless clarity is needed.
\end{remark}

The maximum possible value of $S(\overline{\mathbf{Z}},t)$ for given
$\overline{\mathbf{Z}}$ occurs when $\mathfrak{m}_{k}$ are \emph{equally
probable }(ep):%
\begin{equation}
p_{k}(t)\rightarrow p_{k}^{\text{ep}}=1/W(\overline{\mathbf{Z}}%
)>0,\text{~~\ \ }\forall\mathfrak{m}_{k}\in\Gamma_{0}(\overline{\mathbf{Z}}).
\label{Equiprobable-Z}%
\end{equation}
In this case, the explicit time dependence in $S(t)$ will \emph{disappear} and
we have
\begin{equation}
S_{\text{max}}(\overline{\mathbf{Z}},t)=S(\overline{\mathbf{Z}})=\ln
W(\overline{\mathbf{Z}}), \label{S_Boltzmann0}%
\end{equation}
which is identical in form to the Boltzmann entropy in Eq. (\ref{Boltzmann_S})
for an isolated body in equilibrium, except that the current formulation has
been extended to an isolated body out of equilibrium; see also Eq.
(\ref{Boltzmann_S_Extended}). The only requirement is that all microstates in
$\mathbf{m}_{0}\equiv\mathbf{m}_{0}(\overline{\mathbf{Z}})$ are equally
probable. The statistical entropy in this case becomes a \emph{state
function}, just as the classical entropy is for $\mathcal{M}_{\text{ieq}}$
that was treated in Sect. \ref{Marker_Internal Equilibrium}. As the
equivalence covers $\mathcal{M}_{\text{eq}}$, this proves (3). The proof for
(2) is deferred to Sect. (\ref{Sec-MicrostateProbabilities}), where we discuss
$p_{k}$.

The simplest way to understand the physical meaning of Eq. (\ref{S_Boltzmann0}%
) is as follows: Consider $\overline{\mathbf{Z}}\in\mathfrak{S}_{\mathbf{Z}}$
at some time $t$. As $S(\overline{\mathbf{Z}},t)$ may not be a unique function
of $\overline{\mathbf{Z}}$, we look at all possible entropy functions for this
$\overline{\mathbf{Z}}$. These entropies correspond to all possible sets of
$\left\{  p_{k}(t)\right\}  $ for a fixed $\overline{\mathbf{Z}}$, and define
different possible states $\left\{  \mathcal{M}(\overline{\mathbf{Z}%
})\right\}  $. We pick that particular $\overline{\mathcal{M}}(\overline
{\mathbf{Z}})\in\left\{  \mathcal{M}(\overline{\mathbf{Z}})\right\}  $ among
these that has the \emph{maximum possible value} of the entropy, which we
denote by $S(\overline{\mathbf{Z}})$ or $S(\mathbf{Z}(t))$ without any
explicit $t$-dependence. This entropy is a \emph{state function}
$S(\overline{\mathbf{Z}})$ in $\mathfrak{S}_{\mathbf{Z}}$. For a macroscopic
body, this occurs when the corresponding microstate probabilities for
$\overline{\mathcal{M}}(\overline{\mathbf{Z}})$ are given by $p_{k}%
^{\text{ep}}$ above.

\begin{remark}
\label{Remark-FlatDistribution}We emphasize that $\overline{\mathbf{Z}%
}=(E,\mathbf{W})$ so $p_{k}^{\text{ep}}$ is determined by the average energy
$E$ and not by the microstate energy $E_{k}$ as derived later in Sect.
(\ref{Sec-MicrostateProbabilities}). The equiprobability assumption in Eq.
(\ref{Equiprobable-Z}) basically replaces the actual probability distribution
in Eq. (\ref{microstate probability}) by a \emph{flat distribution} of height
$1/W(\overline{\mathbf{Z}})$ and width $W(\overline{\mathbf{Z}})$, a common
practice in statistical mechanics \cite{Landau}. Despite this modification,
the entropy has the same value $S(\overline{\mathbf{Z}})$\ for a macroscopic
body, for which $\beta$ and $\mathbf{F}_{\text{w}}$ are given by Eq.
(\ref{Fields_Isolated0}).
\end{remark}

Applying the above formulation to a state characterized by a given
$\overline{\mathbf{X}}$ in $\mathfrak{S}_{\mathbf{X}}$\ and consisting of
microstates $\left\{  \overline{m}_{k}\right\}  $, forming the set
$\overline{\mathbf{m}}\equiv\mathbf{m}(\overline{\mathbf{X}}),$ with
probabilities $\left\{  \overline{p}_{k}(t)>0\right\}  $, we find that
\begin{equation}
S(\overline{\mathbf{X}},t)\equiv-%
{\textstyle\sum\limits_{k=1}^{W(\overline{\mathbf{X}})}}
\overline{p}_{k}(t)\ln\overline{p}_{k}(t),%
{\textstyle\sum\limits_{k=1}^{W(\overline{\mathbf{X}})}}
\overline{p}_{k}(t)\equiv1, \label{Conventional_Entropy0}%
\end{equation}
is the entropy of this state, where $W(\overline{\mathbf{X}})$ is the number
of distinct microstates $\overline{m}_{k}$. It should be obvious that%
\[
W(\overline{\mathbf{X}})\equiv%
{\textstyle\sum\nolimits_{\boldsymbol{\xi}(t)}}
W(\overline{\mathbf{Z}}).
\]
Again, under the equiprobable assumption
\[
\overline{p}_{k}(t)\rightarrow\overline{p}_{k}^{\text{ep}}=1/W(\overline
{\mathbf{X}}),\text{~~\ \ }\forall\overline{m}_{k}\in\Gamma(\overline
{\mathbf{X}}),
\]
$\Gamma(\overline{\mathbf{X}})$\ denoting the space spanned by microstates
$\left\{  \overline{m}_{k}\right\}  $, the above entropy takes its maximum
possible value; here,%
\begin{equation}
S_{\text{max}}(\overline{\mathbf{X}},t)=S(\overline{\mathbf{X}})=\ln
W(\overline{\mathbf{X}}), \label{S_Boltzmann}%
\end{equation}
which is identical in value to the Boltzmann entropy in Eq. (\ref{Boltzmann_S}%
) for an isolated body in equilibrium. The maximum value occurs at
$t=\tau_{\text{eq}}$. It is evident that%
\begin{equation}
S(\overline{\mathbf{Z}},t)\leq S(\overline{\mathbf{Z}})\leq S(\overline
{\mathbf{X}}). \label{Entropy_Inequalities0}%
\end{equation}

We will refer to $S(\overline{\mathbf{Z}})$ in terms of microstate number
$W(\overline{\mathbf{Z}})$ in Eq. (\ref{S_Boltzmann0}) as the
\emph{time-dependent Boltzmann formulation }of the entropy or simply the
Boltzmann entropy \cite{Lebowitz}, whereas $S(\overline{\mathbf{X}})$ in Eq.
(\ref{S_Boltzmann}) represents the equilibrium (Boltzmann) entropy. It is
evident that the Gibbs formulation in Eqs. (\ref{Gibbs_Formulation}) and
(\ref{Conventional_Entropy0}) supersedes the Boltzmann formulation in
Eqs.(\ref{S_Boltzmann0}) and (\ref{S_Boltzmann}), respectively, as the former
contains the latter as a special limit. However, it should be also noted that
there are competing views on which entropy is more general
\cite{Lebowitz,Ruelle}. We believe that the above derivation, being general,
makes the Gibbs formulation more fundamental. The continuity of $S(\overline
{\mathbf{Z}},t)$ follows directly from the continuity of $p_{k}(t)$. The
existence of the statistical entropy $S(\overline{\mathbf{Z}},t)$ follows from
the observation that it is bounded above by $\ln W(\overline{\mathbf{Z}})$ and
bounded below by $0$, see Eq. (\ref{S_Boltzmann0}).

It should be stressed that $\mathcal{W}$ is not the number of microstates of
the $\mathcal{N}$ replicas; the latter is given by $[W(\overline{\mathbf{Z}%
})]^{\mathcal{N}}$. Thus, the entropy in Eq.
(\ref{Ensemble_entropy_Formulation}) should not be confused with the Boltzmann
entropy, which would be given by $\mathcal{N}\ln W(\overline{\mathbf{Z}})$. It
should be mentioned at this point that Boltzmann uses the combinatorial
argument to obtain the entropy of a gas, see Eq. (\ref{Boltzmann_S_1}),
resulting in an expression similar to that of the Gibbs entropy in Eq.
(\ref{Gibbs_S}) except that the probabilities appearing in his formulation
represents the probability of various discrete states of a particle, and
should not be confused with the microstate probabilities used here; see Sect.
\ref{Marker_Jaynes}. The approach of Boltzmann is \emph{limited} to that of an
ideal gas only and is not general as it neglects the correlations present due
to the interactions between particles \cite{Jaynes,Lebowitz}. On the other
hand, our approach is valid for any body with any arbitrary interactions
between particles as all microstates in the collection are \emph{independent}.

\subsubsection{System in a Medium and
Quasi-independence\label{marker_Open system}}

The above formulation of $S(\overline{\mathbf{Z}},t)$ can be applied to
$\Sigma,\widetilde{\Sigma}$, and $\Sigma_{0}$. We assume that $\Sigma$, and
$\widetilde{\Sigma}$ are \emph{quasi-independent} so that $S_{0}(t)$ can be
expressed as a sum of entropies $S(t)$ and $\widetilde{S}(t)$ of $\Sigma$, and
$\widetilde{\Sigma}$, respectively:
\begin{equation}
S_{0}(t)=S(t)+\widetilde{S}(t). \label{Entropy Sum}%
\end{equation}
The two statistical entropies are given by an identical formulation
\begin{equation}
S(t)=-%
{\textstyle\sum\limits_{k}}
p_{k}(t)\ln p_{k}(t),\ \ \ \widetilde{S}(t)=-%
{\textstyle\sum\limits_{\widetilde{k}}}
\widetilde{p}_{\widetilde{k}}(t)\ln\widetilde{p}_{\widetilde{k}}(t),
\label{Entropies}%
\end{equation}
respectively. Here, $\mathfrak{m}_{k}$ with probability $p_{k}$ denotes a
microstate of $\Sigma$ and $\widetilde{m}_{\widetilde{k}}$ with probability
$\widetilde{p}_{\widetilde{k}}$ that of the medium. In the derivation of the
above additivity, see \cite{Gujrati-II}, we have neither assumed the medium
nor the system to be in internal equilibrium; only quasi-independence is
assumed. The above formulation of the additivity of statistical entropies will
not remain valid if the two are not quasi-independent. From this, we also
conclude that the additivity will also not be true of the thermodynamic entropies.

\subsection{Disjoint Sample Space (Component
Confinement)\label{Marker_Disjoint Space}}

The consideration of dynamics\ resulting in the simple connectivity of the
sample (or phase) space has played a pivotal role in developing the kinetic
theory of gases \cite{Boltzmann0,Lebowitz}, where the interest is at high
temperatures \cite{Landau,Gujrati-Residual,Gujrati-Symmetry,Gujrati-Poincare}.
As dynamics is very fast here, it is well known that the ensemble entropy
agrees with its temporal formulation. However, at low temperatures, where
dynamics becomes sluggish as in a glass
\cite{Gujrati-book,Palmer,Gutzow,Nemilov}, the body can be \emph{confined}
into disjoint components.

Sample (or phase) space confinement at a phase transition such as a liquid-gas
transition is well known in equilibrium statistical mechanics
\cite{Landau,Gujrati-Residual,Gujrati-Symmetry}. It also occurs when the body
undergoes symmetry breaking such as during magnetic transitions,
crystallizations, etc. But confinement can also occur under nonequilibrium
conditions, when the observational time scale $\tau_{\text{obs}}$\ becomes
shorter than the equilibration time $\tau_{\text{eq}}$
\cite{Gujrati-book,Palmer,Gutzow,Nemilov}, such as for glasses, whose behavior
and properties have been extensively studied. In the following, we will focus
on $\Sigma_{\text{b}}$.

The issue has been recently considered by us \cite{Gujrati-Symmetry}, where
only energy as an observable was considered. The discussion is easily extended
to the present case when confinement occurs for whatever reasons into one of
the thermodynamically significant number of disjoint components $\Gamma
_{\lambda},\lambda=1,2,3\cdots,\mathcal{C}$, each component corresponding to
the same set $\overline{\mathbf{Z}}$. Such a situation arises, for example, in
Ising magnets at the ferromagnetic transition., where the system is either
confined to $\Gamma_{+}$ with positive magnetization or $\Gamma_{-}$ with
negative magnetization. Even a weak external magnetic field $\left\vert
H\right\vert \rightarrow0$, that we can \emph{control} as an observer, will
allow the system to make a choice between the two parts of $\Gamma$. It just
happens that in this case $\mathcal{C}=2$ and is thermodynamically insignificant.

The situation with glasses or other amorphous materials is very different
\cite{Palmer}, In the first place, $\Gamma$\ is a union of
\emph{thermodynamically significant} number $\mathcal{C}\sim e^{cN}%
,c>0,$\ disjoint components. In the second place, there is no analog of a
symmetry breaking field. Therefore, there is no way to prepare a sample in a
given component $\Gamma_{\lambda}$. Thus, the samples will be found in all
different components. Taking into consideration disjointness of the components
generalizes the number of configurations in Eq. (\ref{Combinations}) to%
\[
\mathcal{W\equiv N}!/%
{\textstyle\prod\limits_{\lambda,k_{\lambda}}}
\mathcal{N}_{k_{\lambda}}(t)!,
\]
where $\mathcal{N}_{k_{\lambda}}$ denotes the number of sample in the
microstate $\mathfrak{m}_{k_{\lambda}}$ in the $\lambda$-th component. In
terms of $p_{k_{\lambda}}=\mathcal{N}_{k_{\lambda}}(t)/\mathcal{N}$, this
combination immediately leads to%
\begin{equation}
S(t)\equiv\ln\mathcal{W}/\mathcal{N}=-%
{\textstyle\sum\nolimits_{\lambda}}
{\textstyle\sum\nolimits_{k_{\lambda}}}
p_{k_{\lambda}}(t)\ln p_{k_{\lambda}}(t), \label{S_Component}%
\end{equation}
for the statistical entropy of the system and has already been used earlier
\cite[see Sec. 4.3.3]{Gujrati-Symmetry} by us.\ From what has been said above,
this statistical entropy is also the thermodynamic entropy of a nonequilibrium
state under component confinement for which the entropy is a state function of
$\overline{\mathbf{Z}}$. Therefore, as before, we take $S$ to be the general
expression of the nonequilibrium thermodynamic entropy and use $S$ in place of
$S$.

Introducing%
\[
p_{\lambda}(t)\equiv%
{\textstyle\sum\nolimits_{k_{\lambda}}}
p_{k_{\lambda}}(t),
\]
it is easy to see \cite{Gujrati-Symmetry} that
\[
S(t)=%
{\textstyle\sum\nolimits_{\lambda}}
p_{\lambda}(t)S_{\lambda}(t)+S_{\mathcal{C}}(t).
\]
Here, the entropy of the component $\Gamma_{\lambda}$ in terms of the reduced
microstate probability $\widehat{p}_{k_{\lambda}}\equiv p_{k_{\lambda}%
}/p_{\lambda}$ is%
\begin{equation}
S_{\lambda}(t)=-%
{\textstyle\sum\nolimits_{k_{\lambda}}}
\widehat{p}_{k_{\lambda}}(t)\ln\widehat{p}_{k_{\lambda}}(t)
\label{S_Single_Componet}%
\end{equation}
so that the first contribution is its average over all components. The second
term is given by%
\begin{equation}
S_{\mathcal{C}}(t)=-%
{\textstyle\sum\nolimits_{\lambda}}
p_{\lambda}(t)\ln p_{\lambda}(t), \label{S_Residual}%
\end{equation}
and represents the component entropy. It is this entropy that determines the
residual entropy \cite{Gujrati-Residual} in disordered systems for
$\Sigma_{\text{b}}=\Sigma$. \ 

\subsection{A Proof of The Second Law\label{Sec-SecondLawProof}}

The second law has been proven so far under different assumptions \cite[among
others]{Tolman,Rice,Jaynes,Gujrati-Residual,Gujrati-Symmetry}. Here, we
provide a simple proof of it based on the postulate of the flat distribution;
see Remark \ref{Remark-FlatDistribution}. The current proof is an extension of
the proof given earlier, see \cite[Theorem 4]{Gujrati-Symmetry}. We consider
an isolated system $\Sigma_{\text{b}}=\Sigma_{0}$ for which the second law is
expressed by Eq. (\ref{Second_Law}). As the law requires considering the
instantaneous entropy as a function of time, we need to focus on the sample
space at each instant to determine its entropy $S$ as a function of time. At
each instance, it is an ensemble average over the instantaneous sample space
$\Gamma(t)$ formed by the instantaneous set $\mathbf{m}(t)$ of available
microstates, see Eq. (\ref{Gibbs_Formulation}). We will use the flat
distributions for the microstates at each instance, see Remark
\ref{Remark-FlatDistribution}, so that the entropy is given by Eq.
(\ref{S_Boltzmann0}).

To prove the second law, we proceed in steps by considering a sequence of
sample spaces belonging to $\Gamma(t)$\ as follows
\cite{Gujrati-Residual,Gujrati-Symmetry}. At a given instant, $\Sigma
_{\text{b}}$ happens to be in some microstate. We start at $t=t_{1}=0$, at
which time $\Sigma$ happens to be in a microstate, which we label
$\mathfrak{m}_{1}$. It forms a sample space $\Gamma_{1}$ containing
$\mathfrak{m}_{1}$ with probability $p_{1}^{(1)}=1$, with the superscript
denoting the sample space. We have $S^{(1)}=0$. At some $t=t_{2}$, the sample
space is enlarged from $\Gamma_{1}$ to $\Gamma_{2}$, which contains
$\mathfrak{m}_{1}$ and $\mathfrak{m}_{2}$, with probabilities $p_{1}^{(2)}$
and $p_{2}^{(2)}$. Using the flat distribution, the entropy is now $S_{2}%
=\ln2$. We just follow the system in a sequence of time so that at $t=t_{n}$,
we have a sample space $\Gamma_{n}$ with $\mathfrak{m}_{1},\mathfrak{m}%
_{2},\cdots,\mathfrak{m}_{n}$ so that $S_{n}=\ln n$. Continuing this until all
microstates in $\Gamma$ have appeared, we have $S_{\text{max}}=\ln W$.

Thus, we have proven that the entropy continues to increase until it reaches
its maximum in accordance with the second law in Eq. (\ref{Second_Law}).

\section{Extended State Space, $\mathcal{M}_{\text{ieq}}$ and $\mathcal{M}%
_{\text{nieq}}$\label{Sec-ExtendedS-M-nieq}}

\subsection{Choice of $\mathbf{Z}$ for $\mathcal{M}_{\text{ieq}}%
$\label{Sec-Choice-S_Z}}

We first discuss how to choose a particular state space\ for a unique
description of $\mathcal{M}$ depending on the experimental setup. To
understand the procedure for this, we begin by considering a set
$\boldsymbol{\xi}_{n}$ of internal variables\emph{ }$\left(  \xi_{1},\xi
_{2},\cdots,\xi_{n}\right)  $ and $\mathbf{Z}_{n}\doteq\mathbf{X}%
\cup\boldsymbol{\xi}_{n}$ to form a sequence of state spaces $\mathfrak{S}%
_{\mathbf{Z}}^{(n)}$. In general, one may need many internal variables, with
the value of $n$ increasing as $\mathcal{M}$ is more and more out of EQ
\cite{Gujrati-Hierarchy} relative to $\mathcal{M}_{\text{eq}}$. We will take
$n^{\ast}$ to be the maximum $n$\ in this study as discussed in Sect.
\ref{Sec-InternalVariables}, even though $n$ $<<n^{\ast}$ needed for
$\mathfrak{S}_{\mathbf{Z}}^{(n)}$ will usually be a small number in most
cases. The two most important but distinct time scales are $\tau_{\text{obs}}%
$, the time to make observations, and $\tau_{\text{eq}}$, the equilibration
time for a state $\mathcal{M}$ to turn into $\mathcal{M}_{\text{eq}}$. For
$\tau_{\text{obs}}<\tau_{\text{eq}}$, the system will be in a NEQ state. Let
$\tau_{i}\ $denote the relaxation time of $\xi_{i}$ needed to come to its
equilibrium value so that its affinity $A_{i}\rightarrow0$
\cite{deGroot,Prigogine,Gujrati-Hierarchy,Prigogine71,Gutzow,Nemilov}. For
convenience, we order $\xi_{i}$ so that
\[
\tau_{1}>\tau_{2}>\cdots;
\]
we assume distinct $\tau_{i}$'s for simplicity without affecting our
conclusions. For $\tau_{1}<\tau_{\text{obs}}$, all internal variables have
equilibrated so they play no role in equilibration except thermodynamic forces
$T-T_{0},P-P_{0}$, etc. associated with $\mathbf{X}$ that still drive the
system towards EQ. We choose $n$ satisfying $\tau_{n}>\tau_{\text{obs}}%
>\tau_{n+1}$ so that all of $\xi_{1},\xi_{2},\cdots,\xi_{n}$ have not
equilibrated (their affinities are nonzero). They play an important role in
the NEQT, while $\xi_{n+1},\xi_{n+2},\cdots$ need not be considered as they
have all equilibrated. This specify $\mathcal{M}$ \emph{uniquely} in
$\mathfrak{S}_{\mathbf{Z}}^{(n)}$, which was earlier identified as being in IEQ.

Note that NEQ states with $\tau_{n+1}>\tau_{\text{obs}}>\tau_{n+2}$ are not
uniquely identifiable in $\mathfrak{S}_{\mathbf{Z}}^{(n)}$, even though they
are uniquely identifiable in $\mathfrak{S}_{\mathbf{Z}}^{(n+1)}$. Thus, there
are many NEQ states that are not unique in $\mathfrak{S}_{\mathbf{Z}}^{(n)}$.
The unique states $\mathcal{M}_{\text{ieq}}$\ are special in that its Gibbs
entropy $S(\mathbf{Z}_{n})$\ is a state function of $\mathbf{Z}_{n}$ in
$\mathfrak{S}_{\mathbf{Z}}^{(n)}$. Thus, given $\tau_{\text{obs}}$, we look
for the window $\tau_{n}>\tau_{\text{obs}}>\tau_{n+1}$ to choose the
particular value of $n$. This then determines $\mathfrak{S}_{\mathbf{Z}}%
^{(n)}$ in which the states are in IEQ. From now onward, we assume that $n$
has been found and $\mathfrak{S}_{\mathbf{Z}}^{(n)}$ has been identified. We
now suppress $n$ and simply use $\mathfrak{S}_{\mathbf{Z}}$ below.

\subsection{Microstate probabilities for $\mathcal{M}_{\text{ieq}}%
$\label{Sec-MicrostateProbabilities}}

As $\mathcal{M}_{\text{ieq}}$ is unique in $\mathfrak{S}_{\mathbf{Z}}$, we
need to identify the unique set $\left\{  p_{k}\right\}  $. As we keep
$\mathbf{W}$ fixed in $\mathcal{M}_{\text{ieq}}$ as the parameter, then
$\mathbf{F}_{\text{w}k}$, the value $\mathbf{F}_{\text{w}}$ takes on
$\mathfrak{m}_{k}$,\ are fluctuating forces in $\mathfrak{S}_{\mathbf{Z}}$
that satisfy a sum rule, just as the microstate energies $E_{k}$:
\[
E=\sum_{k}E_{k}p_{k},\mathbf{F}_{\text{w}}=\sum_{k}\mathbf{F}_{\text{w}k}%
p_{k}.
\]
We need to maximize the entropy $S(\mathbf{Z})$ at fixed $E$ and
$\mathbf{F}_{\text{w}}\ $by varying $p_{k}$ without changing $\left\{
\mathfrak{m}_{k}\right\}  $ , i.e. without changing $E_{k}$ and $\mathbf{F}%
_{\text{w}k}$ to identify the particular set $\left\{  p_{k}\right\}  $ to
achieve it. Using the Lagrange multiplier technique, it is easy to show that
the condition for this in terms of three Lagrange multipliers with obvious
definitions is
\begin{equation}
\eta_{k}=\lambda_{1}+\lambda_{2}E_{k}+\boldsymbol{\lambda}_{3}\cdot
\mathbf{F}_{\text{w}k}, \label{index_i}%
\end{equation}
from which follows the statistical entropy $\mathcal{S}=-(\lambda_{1}%
+\lambda_{2}E+\boldsymbol{\lambda}_{3}\cdot\mathbf{F}_{\text{w}})$; we have
reverted back to the original symbol for the statistical entropy here. It is
now easy to identify $\lambda_{2}=-\beta,\boldsymbol{\lambda}_{3}%
\beta\mathbf{W}$ by comparing $d\mathcal{S}$ with $dS$ in Eq.
(\ref{Gibbs_Fundamental_Extended}) by varying $E$ and $\mathbf{W}$ so we
finally have
\begin{equation}
p_{k}=\exp[\beta(\widehat{G}-E_{k}-\mathbf{W}\cdot\mathbf{F}_{\text{w}k})],
\label{microstate probability}%
\end{equation}
where $\lambda_{1}=\beta\widehat{G}$\ with $\widehat{G}(t)$ is a normalization
constant and defines a NEQ partition function
\begin{subequations}
\begin{equation}
\exp(-\beta\widehat{G})\equiv\sum_{k}\exp[-\beta(E_{k}+\mathbf{W}%
\cdot\mathbf{F}_{\text{w}k})]. \label{NEQ-PF}%
\end{equation}
It is easy to verify that
\begin{equation}
\widehat{G}(T,\mathbf{W})=E+\mathbf{W}\cdot\mathbf{F}_{\text{w}}-TS,
\label{NEQ-Potential}%
\end{equation}
so that if we neglect the fluctuations $E_{k}-E$ and $\mathbf{F}_{\text{w}%
k}-\mathbf{F}_{\text{w}}$, then $p_{k}$ reduces to the flat distribution
$p_{k}=1/W(E,\mathbf{W})$ in Remark \ref{Remark-FlatDistribution}, which can
be identified as the microstate probability in the NEQ microcanonical ensemble.

It should be remarked that the Lagrange multipliers in $p_{k}$ are determined
by comparing $d\mathcal{S}$ with $dS$ in the Gibbs fundamental relation, a
thermodynamic relation. This then proves that $\mathcal{S}$ is the same as the
thermodynamic entropy $S$ up to a constant \cite{Gujrati-Entropy2}, which can
be fixed by appeals to the third law. We do not consider here the issue of a
residual entropy, which is discussed elsewhere
\cite{Gujrati-Residual,Gujrati-Hierarchy} and can be done based on the
discussion in Sect. \ref{Marker_Disjoint Space}. The $p_{k}$ above clearly
shows the effect of irreversibility and is very different from its equilibrium
analog $p_{k}^{\text{eq}}$ in $\mathfrak{S}_{\mathbf{X}}$:
\end{subequations}
\begin{equation}
p_{k}^{\text{eq}}=\exp[\beta_{0}(\widehat{G}(T_{0},\mathbf{w})-E_{k}%
-\mathbf{w}\cdot\mathbf{f}_{\text{w}0k})] \label{microstate probability-EQ}%
\end{equation}
obtained by replacing $\mathbf{W}$ by $\mathbf{w}$, $\mathbf{F}_{\text{w}k}$
by $\mathbf{f}_{\text{w}0k}$,\ and $\beta$ by $\beta_{0}$. The fluctuating
$E_{k},\mathbf{f}_{\text{w}k}$ satisfy%
\[
E=\sum_{k}E_{k}p_{k}^{\text{eq}},\mathbf{f}_{\text{w}0}=\sum_{k}%
\mathbf{f}_{\text{w}0k}p_{k}^{\text{eq}}.
\]

The observation time $\tau_{\text{obs}}$ is determined by the way $T$ and
$\mathbf{W}$ are changed during a process. Thus, during each change,
$\tau_{\text{obs}}$ must be compared with the time needed for $\Sigma$ to come
to the next IEQ state, and for the microstate probabilities to be given by Eq.
(\ref{microstate probability}) with the new values of $T$ and $\mathbf{W}$.

\subsection{$\mathcal{M}_{\text{nieq}}$ in $\mathfrak{S}_{\mathbf{Z}}$
\label{Sec-M_nieq}}

We now focus on a non-unique state $\mathcal{M}_{\text{nieq}}$ in
$\mathfrak{S}_{\mathbf{Z}}$. This will be needed if $\tau_{\text{obs}}$ is
reduced to make the process faster so that instead of falling in the window
($\tau_{n},\tau_{n+1}$), it now falls in a higher window such as ($\tau
_{n+1},\tau_{n+2}$)$.$ As said above, $\mathcal{M}$ can now be treated as a
unique state in a larger state space $\mathfrak{S}_{\mathbf{Z}^{\prime}%
}\supset\mathfrak{S}_{\mathbf{Z}}$. Let $\boldsymbol{\xi}^{\prime}(t)$ denote
the set of additional internal variables needed over $\mathfrak{S}%
_{\mathbf{Z}}$ so that
\[
\mathbf{Z}^{\prime}(t)=(\mathbf{Z}(t),\boldsymbol{\xi}^{\prime}(t)).
\]
The entropy $S(\mathbf{Z}^{\prime}(t))=S(\mathbf{Z}(t),t)$ for $\mathcal{M}%
_{\text{ieq}}(t)$\ in $\mathfrak{S}_{\mathbf{Z}^{\prime}}$ satisfies the Gibbs
fundamental relation%
\begin{subequations}
\begin{equation}
dS(\mathbf{Z}^{\prime}(t))\mathbf{=}\frac{\partial S}{\partial E}%
dE\mathbf{+}\frac{\partial S}{\partial\mathbf{W}}\cdot d\mathbf{W+}%
\frac{\partial S}{\partial\boldsymbol{\xi}^{\prime}}\cdot d\boldsymbol{\xi
}^{\prime}, \label{GibbsFR-S-NIEQ-prime}%
\end{equation}
where $\mathbf{W}$ is the work variable in $\mathfrak{S}_{\mathbf{Z}}$.
Expressing the last term as%
\end{subequations}
\[
\frac{\partial S}{\partial\boldsymbol{\xi}^{\prime}}\cdot\frac
{d\boldsymbol{\xi}^{\prime}}{dt}dt,
\]
we obtain the following generalization of the Gibbs fundamental relation for
$\mathcal{M}_{\text{nieq}}(t)$\ in $\mathfrak{S}_{\mathbf{Z}}$:%
\begin{equation}
dS(\mathbf{Z}(t),t)=\frac{\partial S}{\partial E}dE\mathbf{+}\frac{\partial
S}{\partial\mathbf{W}}\cdot d\mathbf{W+}\frac{\partial S}{\partial t}dt,
\label{GibbsFR-S-NIEQ}%
\end{equation}
where%
\begin{equation}
\frac{\partial S}{\partial t}\doteq\frac{\partial S}{\partial\boldsymbol{\xi
}^{\prime}}\cdot\frac{d\boldsymbol{\xi}^{\prime}}{dt}\geq0. \label{dS/dt-NIEQ}%
\end{equation}
In $\mathfrak{S}_{\mathbf{Z}^{\prime}}$, we can identify the temperature $T$
as the thermodynamic temperature in $\mathfrak{S}_{\mathbf{Z}^{\prime}}$ using
the standard definition. But, it is clear from the above discussion that
$\partial S(\mathbf{Z}^{\prime}(t))/\partial E$ in $\mathfrak{S}%
_{\mathbf{Z}^{\prime}}$ has the same value as $\partial S(\mathbf{Z}%
(t),t)/\partial E$ in $\mathfrak{S}_{\mathbf{Z}}$. Therefore, we are now set
to identify $T$ as a thermodynamic temperature for any arbitrary state
$\mathcal{M}$.

\begin{remark}
The temperature $T$ in $\mathfrak{S}_{\mathbf{Z}}$ and $\mathfrak{S}%
_{\mathbf{Z}^{\prime}}$ are the same
\begin{equation}
\beta=1/T=\partial S(\mathbf{Z}^{\prime}(t))/\partial E=\partial
S(\mathbf{Z}(t),t)/\partial E. \label{beta_arb}%
\end{equation}

\end{remark}

\begin{definition}
\label{Def-HiddenEntropy}As the presence of $\partial S/\partial t$ above in
$\mathfrak{S}_{\mathbf{Z}}$ is due to "hidden" internal variables in
$\boldsymbol{\xi}^{\prime}$, we will call it the \emph{hidden entropy
generation rate}, and
\begin{subequations}
\begin{equation}
d_{\text{i}}S^{\text{hid}}(t)=\frac{\partial S}{\partial t}dt=\frac{\partial
S}{\partial\boldsymbol{\xi}^{\prime}}\cdot d\boldsymbol{\xi}^{\prime}\geq0,
\label{hiddenn-entropy-generation}%
\end{equation}
the \emph{hidden entropy generation}. It results in a \emph{hidden
irreversible work}%
\begin{equation}
d_{\text{i}}W^{\text{hid}}\doteq Td_{\text{i}}S^{\text{hid}}=\mathbf{A}%
^{\prime}\cdot d\boldsymbol{\xi}^{\prime}, \label{hiddenn-irreversible-work}%
\end{equation}
in $\mathfrak{S}_{\mathbf{Z}}$ due to the hidden internal variable with
affinity $\mathbf{A}^{\prime}$.
\end{subequations}
\end{definition}

\begin{remark}
\label{Remark-IEQ-ARB-Macrostate}A state $\mathcal{M}_{\text{nieq}}(t)$ with
$S(\mathbf{Z}(t),t)$ can be converted to $\mathcal{M}_{\text{ieq}}(t)$ with a
state function $S(\mathbf{Z}^{\prime}(t))$ in an appropriately chosen state
space $\mathfrak{S}_{\mathbf{Z}^{\prime}}\supset\mathfrak{S}_{\mathbf{Z}}$ by
finding the appropriate window in which $\tau_{\text{obs}}$ lies. The needed
additional internal variable $\boldsymbol{\xi}^{\prime}$\ determines the
hidden entropy generation rate $\partial S/\partial t$ in Eq.
(\ref{dS/dt-NIEQ}) due to the non-IEQ nature of $\mathcal{M}_{\text{nieq}}%
(t)$\ in $\mathfrak{S}_{\mathbf{Z}}$, and ensures validity of the Gibbs
relation in Eq. (\ref{GibbsFR-S-NIEQ}) for it, thereby providing not only a
new interpretation of the temporal variation of the entropy due to hidden
variables but also extends the MNEQT to $\mathcal{M}_{\text{nieq}}(t)$\ in
$\mathfrak{S}_{\mathbf{Z}}$.
\end{remark}

The above discussion strongly points towards the possible

\begin{proposition}
\label{Proposition-General-MNEQT}The MNEQT provides a very general framework
to study any $\mathcal{M}_{\text{nieq}}(t)$ in $\mathfrak{S}_{\mathbf{Z}}$,
since it can be converted into a $\mathcal{M}_{\text{ieq}}(t)$ in an
appropriately chosen state space $\mathfrak{S}_{\mathbf{Z}^{\prime}}$, with
$d_{\text{i}}S^{\text{hid}}(t)$ originating from hidden internal variable
$\boldsymbol{\xi}^{\prime}$.
\end{proposition}

\begin{remark}
In a process $\mathcal{P}$ resulting in $\mathcal{M}_{\text{nieq}}(t)$ in
$\mathfrak{S}_{\mathbf{Z}}$, it is natural to assume that the terminal states
in $\mathcal{P}$ are $\mathcal{M}_{\text{ieq}}$ so the affinity corresponding
to $\boldsymbol{\xi}^{\prime}$ must vanish in them.
\end{remark}

\begin{remark}
\label{Remark-NonIEQ-S_X} By replacing $\mathbf{Z}$ by $\mathbf{X}$, and
$\mathbf{Z}^{\prime}$ by $\mathbf{Z}$, we can also express the Gibbs
fundamental relation for any NEQ state in $\mathfrak{S}_{\mathbf{X}}$ as%
\begin{equation}
dS(\mathbf{X}(t),t)=\frac{\partial S}{\partial E}dE\mathbf{+}\frac{\partial
S}{\partial\mathbf{w}}\cdot d\mathbf{w+}\frac{\partial S}{\partial t}dt,
\label{Gibbs-FR-NonIEQ-S_X}%
\end{equation}
by treating $\mathcal{M}$ as $\mathcal{M}_{\text{ieq}}$ in $\mathfrak{S}%
_{\mathbf{Z}}$. In a NEQ process $\overline{\mathcal{P}}$ between two
EQ\ states but resulting in $\mathcal{M}_{\text{nieq}}(t)$ between them in
$\mathfrak{S}_{\mathbf{Z}}$, the affinity corresponding to $\boldsymbol{\xi}$
must vanish in the terminal EQ states of $\overline{\mathcal{P}}$.
\end{remark}

Eq. (\ref{Gibbs-FR-NonIEQ-S_X}) proves extremely useful to describe
$\mathcal{M}$ in $\mathfrak{S}_{\mathbf{X}}$ as it may not be easy to identify
$\boldsymbol{\xi}$ in all cases.

\begin{remark}
\label{Remark-dQ-dS-GeneralRelation} The explicit time dependence in the
entropy for $\mathcal{M}_{\text{neq}}$ in $\mathfrak{S}_{\mathbf{X}}$ or
$\mathcal{M}_{\text{ieq}}(t)$\ in $\mathfrak{S}_{\mathbf{Z}}$ is solely due to
the internal variables, which do not affect $dQ=TdS$, with $T$ defined as the
inverse of $\partial S/\partial E$ at fixed $\mathbf{w},t$\ or $\mathbf{W},t$
in the two state spaces, respectively.
\end{remark}

\section{A Model Entropy Calculation\label{Sec-EntropyCalculation}}

We consider a gas of non-interacting identical structureless particles with no
spin, each of mass $m$, in a fixed region confined by impenetrable walls
(infinite potential well). Initially, the gas is in a NEQ state, and is
isolated in that region. In time, the gas will equilibrate and the microstate
probabilities change in a way that the entropy increases. We wish to
understand how the increase happens.

\subsection{1-dimensional ideal Gas:\ }

In order to be able to carry out an \emph{exact calculation}, we consider the
gas in a $1$-dimensional box of initial size $L_{\text{in}}$. As there are no
interactions between the particles, the wavefunction $\Psi$ for the gas is a
product of individual particle wavefunctions $\psi$. Thus, we can focus on a
single particle to study the nonequilibrium behavior of the gas
\cite{Gujrati-JensenInequality,Gujrati-QuantumHeat,GujTyler,GujBoyko,Wu}. The
simple model of a particle in a box has been extensively studied in the
literature but with a very different emphasis \cite{Bender,Doescher,Stutz}.
The particle only has non-degenerate eigenstates whose energies are determined
by $L$, and \ a quantum number $k$. We use the energy scale $\varepsilon
_{1}=\pi^{2}\hbar^{2}/2mL^{2}$ to measure the energy of the eigenstate, and
$\alpha=L/L_{\text{in}}$ so that%
\begin{equation}
\varepsilon_{k}(L)=k^{2}/\alpha^{2}; \label{particle-microstate energy}%
\end{equation}
the corresponding eigenfunction is given by%
\begin{equation}
\psi_{k}(x)=\sqrt{2/L}\sin(k\pi x/L),\ \ k=1,2,3,\cdots.
\label{eigenfunctions}%
\end{equation}
The pressure generated by the eigenstate on the walls is given by
\cite{Landau-QM}
\begin{equation}
P_{k}(L)\equiv-\partial\varepsilon_{k}/\partial L=2\varepsilon_{k}(L)/L.
\label{particle-microstate presure}%
\end{equation}
In terms of the eigenstate probability $p_{k}(t)$, the average energy and
pressure are given by
\begin{subequations}
\label{particle energy pressure}%
\begin{align}
\varepsilon(t,L)  &  \equiv%
{\textstyle\sum\nolimits_{k}}
p_{k}(t)\varepsilon_{k}(L),\label{particle energy}\\
P(t,L)  &  \equiv%
{\textstyle\sum\nolimits_{k}}
p_{k}(t)P_{k}(L)=2\varepsilon(t,L)/L. \label{particle pressure0}%
\end{align}
The single particle entropy follows from Eq. (\ref{Gibbs_Formulation}) by
using the single particle probability $p_{k}(t)$. The time dependence in
$\varepsilon(t)$\ or $P(t)$ is due to the time dependence in $p_{k}$ and
$\varepsilon_{k}(L)$. Even for an isolated system, for which $\varepsilon$
remains constant, $p_{k}$ cannot remain constant as follows directly from the
second law \cite{Gujrati-Entropy1} and creates a conceptual problem because
the eigenstates are mutually orthogonal and there can be no transitions among
them to allow for a change in $p_{k}$.\ 

As the gas is isolated, its energy, volume and the number of particles remain
constant. As it is originally not in equilibrium, it will eventually reach
equilibrium in which its entropy must increase. This requires the introduction
of some internal variables even in this system whose variation will give rise
to entropy generation by causing internal variations $d_{\text{i}}p_{k}(t)$ in
$p_{k}(t)$. Here, we will assume a single internal variable $\xi(t)$. What is
relevant is that the variation in $\xi(t)$ is accompanied by changes
$dp_{k}(t)$ occurring within the isolated system.

\subsection{Chemical Reaction Approach}

A way to change $p_{k}$ in an isolated system is to require the presence of
some stochastic interactions, whose presence allows for transitions among
eigenstates \cite{Gujrati-Symmetry}. As these transitions are happening within
the system, we can treat them as "chemical reactions" between different
eigenstates \cite{DeDonder,deGroot,Prigogine} by treating each eigenstate $k$
as a chemical species. During the transition, these species undergo chemical
reactions to allow for the changes in their probabilities.

We follow this analogy further and extend the traditional approach
\cite{DeDonder,deGroot,Prigogine} to the present case. For the sake of
simplicity, our discussion will be limited to the ideal gas in a box; the
extension to any general system is trivial. Therefore, we will use microstates
$\left\{  \mathfrak{m}_{k}\right\}  $ instead of eigenstates in the following
to keep the discussion general. Let there be $N_{k}(t)$ particles in
$\mathfrak{m}_{k}$ at some instant $t$ so that
\end{subequations}
\[
N=%
{\textstyle\sum\nolimits_{k}}
N_{k}(t)
\]
at all times, and $p_{k}(t)=N_{k}(t)/N$. We will consider the general case
that also includes the case in which final microstates refer to a box size
$L^{\prime}$ different from its initial value $L$. Let us use $A_{k}$ to
denote the reactants (initial microstates) and $A_{k}^{\prime}$ to denote the
products (final microstates). For the sake of simplicity of argument, we will
assume that transitions between microstates is described by a single chemical
reaction, which is expressed in stoichiometry form as
\begin{equation}%
{\textstyle\sum\nolimits_{k}}
a_{k}A_{k}\longrightarrow%
{\textstyle\sum\nolimits_{k}}
a_{k}^{\prime}A_{k}^{\prime}. \label{General Reaction}%
\end{equation}
Let $N_{k}$ and $N_{k}^{\prime}$ denote the population of $A_{k}$ and
$A_{k}^{\prime}$, respectively, so that $N=%
{\textstyle\sum\nolimits_{k}}
N_{k}=%
{\textstyle\sum\nolimits_{k}}
N_{k}^{\prime}$. Accordingly, $p_{k}(t)=N_{k}(t)/N$ for the reactant and
$p_{k}(t+dt)=N_{k}^{\prime}(t)/N$ for the product. The single reaction is
described by a single extent of reaction $\xi$ and we have
\[
d\xi(t)\equiv-dN_{k}(t)/a_{k}(t)\equiv dN_{k^{\prime}}^{\prime}%
(t)/a_{k^{\prime}}^{\prime}(t)\text{ \ \ for all }k,k^{\prime}.
\]
It is easy to see that the coefficients satisfy an important relation%
\[%
{\textstyle\sum\nolimits_{k}}
a_{k}(t)=%
{\textstyle\sum\nolimits_{k}}
a_{k}^{\prime}(t),
\]
which reflects the fact that the change $\left\vert dN\right\vert $ in the
reactant microstates is the same as in the product microstates. The
\emph{affinity} in terms of the chemical potentials $\mu$ is given by%
\[
A(t)=%
{\textstyle\sum}
a_{k}(t)\mu_{A_{k}}(t)-%
{\textstyle\sum}
a_{k}^{\prime}(t)\mu_{A_{k}^{\prime}}(t),
\]
and will vanish only in "equilibrium," i.e. only when $p_{k}$' s attain their
equilibrium values. Otherwise, $A(t)$ will remain non-zero. It acts as the
thermodynamic force in driving the chemical reaction
\cite{DeDonder,deGroot,Prigogine}. But we must wait long enough for the
reaction to come to completion, which happens when $A(t)$ and $d\xi/dt$ both
vanish. The extent of reaction $\xi$ is an example of an internal variable.
There may be other internal variables depending on the initial NEQ state as
discussed in Sect. \ref{Sec-InternalVariables}.

\section{Simple Applications\label{Sec-Applications}}

\subsection{Composite $\Sigma$ with Temperature Inhomogeneity.
\label{Sec-Composite System}}

Here, we will show by examples that the thermodynamic temperature $T$ of
$\Sigma$ allows us to treat it as a "black box" $\Sigma_{\text{B}}$ without
knowing its detailed internal structure such as its composition in terms of
two subsystems $\Sigma_{1}$ and $\Sigma_{2}$. Alternatively, we can treat
$\Sigma$ as a combination $\Sigma_{\text{C}}$ of $\Sigma_{1}$ at $T_{1}$\ and
$\Sigma_{2}$ at at $T_{2}<T_{1}$, and obtain same thermodynamics. Thus, both
approaches are equivalent, which justifies the usefulness of $T$ as
thermodynamically appropriate global temperature.

In the following, we will consider various cases that can be obtained as
special cases of the following general situation: $\Sigma_{1}$ in thermal
contact with a heat medium $\widetilde{\Sigma}_{\text{h1}}$ at temperature
$T_{01}$, and $\Sigma_{2}$ in thermal contact with another heat medium
$\widetilde{\Sigma}_{\text{h2}}$ at temperature $T_{02}$, with the two media
having no mutual interaction.

We will consider the two realizations for $\Sigma$: $\Sigma_{\text{B}}$ and
$\Sigma_{\text{C}}$ to compare their predictions. As discussed for the case
(b) in Sect. \ref{Sec-InternalVariables}, $\Sigma_{1}$ and $\Sigma_{2}$ are
always taken to be in EQ, but $\Sigma$ in IEQ. The entropies in the two
realizations are%
\[
S_{\text{B}}(t)=S(E(t),\xi(t));S_{\text{C}}=S_{1}(E_{1}(t))+S_{2}(E_{2}(t)),
\]
and have the same value; recall that $E(t)=E_{1}(t)+E_{2}(t)$, and
$\xi(t)=E_{1}(t)-E_{2}(t)$ for $\Sigma(t)$; see Eq. (\ref{Internal Variable-1}%
). For clarity, we will often use the argument $t$ to emphasize the variations
in time $t$ in this section. In general, the irreversible entropy generation
is given by%
\begin{equation}
d_{\text{i}}S(t)=d\widetilde{S}_{1}(t)+d\widetilde{S}_{2}(t)+dS(t),
\label{Irreversible EntropyGeneration-Inhomogeneity}%
\end{equation}
where $dS$ should be replaced by $dS_{\text{B}}$ or $dS_{\text{C}}$ as the
case may be:%
\begin{equation}%
\begin{tabular}
[c]{l}%
$dS_{\text{B}}(t)=\beta(t)dE(t)+\beta(t)A(t)d\xi(t),$\\
$dS_{\text{C}}(t)=\beta_{1}(t)dE_{1}(t)+\beta_{2}(t)dE_{2}(t),$%
\end{tabular}
\ \ \ \ \ \label{dSbc}%
\end{equation}
where we are using the inverse temperatures for various bodies. Let
$d_{\text{e}}Q_{l}(t),l=1,2$ be the energy or heat transferred to $\Sigma
_{l}(t)$ from $\widetilde{\Sigma}_{\text{h}}^{(l)}$, and $dE_{\text{in}%
}(t)=d_{\text{e}}Q_{\text{in}}(t)$ the energy or heat transferred from
$\Sigma_{1}(t)$ to $\Sigma_{2}(t)$. We have, using $\delta_{1}=-1$ and
$\delta_{2}=+1,$
\begin{subequations}
\begin{align}
dE_{l}(t)  &  =d_{\text{e}}Q_{l}(t)+\delta_{l}dE_{\text{in}}(t),\nonumber\\
dE(t)  &  =d_{\text{e}}Q_{1}(t)+d_{\text{e}}Q_{2}(t),\label{dEl-dE-dSl}\\
d\widetilde{S}_{l}(t)  &  =-d_{\text{e}}S_{l}(t)=-\beta_{0l}d_{\text{e}}%
Q_{l}(t).\nonumber
\end{align}
We see that $dE(t)$ is unaffected by the internal energy transfer
$dE_{\text{in}}(t)$, while
\begin{equation}
d\xi(t)=d_{\text{e}}Q_{1}(t)-d_{\text{e}}Q_{2}(t)+2dE_{\text{in}}(t),
\label{General-dXi}%
\end{equation}
is affected by the heat exchange disparity $d_{\text{e}}Q_{1}(t)-d_{\text{e}%
}Q_{2}(t)$ along with $dE_{\text{in}}(t)$.

We finally have%
\end{subequations}
\begin{equation}
d_{\text{i}}S(t)=-%
{\textstyle\sum\nolimits_{l}}
\beta_{0l}d_{\text{e}}Q_{l}(t)+dS. \label{diS-Media}%
\end{equation}
We now consider various cases to make an important point.

\subsubsection{Isolated $\Sigma$}

We first consider the realizations $\Sigma_{\text{B}}$. Using $dE(t)=dE_{1}%
(t)+dE_{2}(t),d\xi(t)=dE_{1}(t)-dE_{2}(t)$, see Eqs.
(\ref{Internal Variable-1}), and (\ref{dSbc}) for $dS_{\text{B}}(t)$ above, we
obtain
\begin{subequations}
\begin{equation}
\beta(t)=\frac{\beta_{1}(t)+\beta_{2}(t)}{2},\beta(t)A(t)=\frac{\beta
_{1}(t)-\beta_{2}(t)}{2}. \label{beta-A-Composite}%
\end{equation}
This identifies $T\left(  t\right)  $ in terms of $T_{1}(t)$ and $T_{2}(t)$.
As EQ is attained, $T(t)\rightarrow T_{0}$, the EQ temperature between
$\Sigma_{1}$ and $\Sigma_{2}$, and $A(t)\rightarrow A_{0}=0$ as expected. In
the following, we will use $A^{\prime}(t)$ for $\beta(t)A(t)$ for simplicity.
In terms of $\beta$ and $A^{\prime}$, we also have%
\begin{equation}
\beta_{1}=\beta+A^{\prime},\beta_{2}=\beta-A^{\prime}. \label{beta1-beta2}%
\end{equation}

We now justify that in this simple example, $A^{\prime}(t)d\xi(t)$ determines
$d_{\text{i}}S(t)$ due to irreversibilty in $\Sigma(t)$ for which $dQ=TdS$
reduces to $d_{\text{i}}Q=Td_{\text{i}}S$. Setting $dE(t)=0$ in $dS_{\text{B}%
}(t)$, we have by direct evaluation,
\end{subequations}
\begin{equation}
d_{\text{i}}S(t)=A^{\prime}(t)d\xi(t)=\beta(t)d_{\text{i}}W(t),
\label{diS-Isolated-Composite}%
\end{equation}
where we have also used Eq. (\ref{diQ-diW}). It should be emphasized that the
existence of $d_{\text{i}}S(t)\geq0$ due to $\xi$ in $\mathcal{M}_{\text{ieq}%
}$ is consistent with $\mathcal{M}_{\text{ieq}}$ as a NEQ state, even though
its entropy is a state function in the extended state space.

We now consider $\Sigma_{\text{C}}$, which is also very instructive to
understand the origin of $d_{\text{i}}S(t)$ in a different way. Considering
internal energy or heat transfer $dE_{\text{in}}(t)=d_{\text{e}}Q_{\text{in}%
}(t)$ between $\Sigma_{1}(t)$ and $\Sigma_{2}(t)$ at some instant $t$, we
have
\begin{subequations}
\begin{equation}
dS_{1}(t)=\frac{dE_{\text{in}}(t)}{T_{1}(t)},dS_{2}(t)=-\frac{dE_{\text{in}%
}(t)}{T_{2}(t)}, \label{dS1-dS2-dEin}%
\end{equation}
due to this transfer. This results in
\begin{equation}
d_{\text{i}}S(t)=\left[  \beta_{1}(t)-\beta_{2}(t)\right]  dE_{\text{in}%
}(t)=A^{\prime}d\xi(t), \label{diS-dEin}%
\end{equation}
since $d\xi(t)=dE_{1}(t)-dE_{2}(t)=2dE_{\text{in}}(t)$. Thus, the physical
origin of $d_{\text{i}}S(t)$ is the internal entropy change of the subsystems,
and shows how $d_{\text{i}}S(t)$ can be measured by measuring the EQ
temperatures $T_{1}$\ and $T_{2}$ and $dE_{\text{in}}(t)$ between them. It is
this internal energy flow that gives rise to $\xi$ in this case.

\subsubsection{$\Sigma$ Interacting with $\widetilde{\Sigma}_{\text{h}}$}

To further appreciate the physical significance of the NEQ $T(t)$ of the above
composite system $\Sigma(t)$, we allow it to interact with a heat medium
$\widetilde{\Sigma}_{\text{h}}$ at its EQ temperature $T_{0}$. For this, we
take $\widetilde{\Sigma}_{\text{h1}}$ and $\widetilde{\Sigma}_{\text{h2}}$ at
the same common temperature $T_{0}=T_{01}=T_{02}$ above so that we can treat
them as a single medium $\widetilde{\Sigma}_{\text{h}}$ with heat exchange
$d_{\text{e}}Q(t)$. We thus obtain from Eq. (\ref{diS-Media})
\end{subequations}
\[
d_{\text{i}}S(t)=-\beta_{0}d_{\text{e}}Q(t)+dS.
\]
We will consider two different kinds of interaction below:

\qquad(i) We first consider $\Sigma_{\text{B}}(t)$ in $\mathcal{M}%
_{\text{ieq}}$ at $T(t)$ so we use $dS_{\text{B}}(t)$ above. We thus have
\begin{equation}
d_{\text{i}}S(t)=[\beta(t)-\beta_{0}]d_{\text{e}}Q(t)+A^{\prime}(t)d\xi(t),
\label{Single-Interacting diS}%
\end{equation}
which is consistent with the general identity
\[
d_{\text{i}}W=TdS-T_{0}d_{\text{i}}S=(T-T_{0})d_{\text{e}}S+Td_{\text{i}}S
\]
where $d_{\text{i}}W$ is replaced using Eq. (\ref{diS-Isolated-Composite}).
The above identity is derived for a single (composite)\ system at temperature
$T\left(  t\right)  $. This confirms that the composite $\Sigma_{\text{C}}$
here can be treated as a noncomposite $\Sigma_{\text{B}}$\ at $T\left(
t\right)  $. To be convinced that the above $d_{\text{i}}S(t)$ includes the
internally generated irreversibility in Eq. (\ref{diS-Isolated-Composite}) due
to heat transfer between $\Sigma_{1}(t)$ and $\Sigma_{2}(t)$, we only have to
set $d_{\text{e}}S(t)=0$ to ensure the isolation of $\Sigma$. We reproduce Eq.
(\ref{diS-Isolated-Composite}) as $d_{\text{i}}Q(t)=d_{\text{i}}W(t)$. The
remaining source of irreversibility $T(t)d_{\text{i}}S^{\text{Q}}(t)$ given by
the first term above is due to external heat exchange\ between $\Sigma$ and
$\widetilde{\Sigma}_{\text{h}}$%
\begin{subequations}
\begin{equation}
d_{\text{i}}S^{\text{Q}}(t)=[T_{0}\beta(t)-1]d_{\text{e}}S(t),
\label{diS_Sigma-s}%
\end{equation}
as expected.

\qquad(ii) We take treat $\Sigma(t)$ as $\Sigma_{\text{C}}(t)$ in contact with
$\widetilde{\Sigma}_{\text{h}}$. We deal directly with the two heat exchanges
$d_{\text{e}}Q_{l}(t),l=1,2$ to $\Sigma_{l}(t)$ from $\widetilde{\Sigma
}_{\text{h}}$, and the internal energy transfer $dE_{\text{in}}(t)$. Using
$dE_{l}(t)$ from Eq. (\ref{dEl-dE-dSl}) in $dS_{\text{C}}$ given in Eq.
(\ref{diS-Media}), we find that%
\end{subequations}
\[
d_{\text{i}}S(t)=%
{\textstyle\sum\nolimits_{l}}
[\beta_{l}(t)-\beta_{0}]d_{\text{e}}Q_{l}(t)+\left[  \beta_{1}(t)-\beta
_{2}(t)\right]  dE_{\text{in}}(t).
\]
Using Eq. (\ref{beta1-beta2}) to express $\beta_{l}$, we can rewrite the above
equation as%
\[
d_{\text{i}}S(t)=[\beta(t)-\beta_{0}]d_{\text{e}}Q(t)+A^{\prime}d\xi,
\]
where we have used the identity
\begin{equation}
d_{\text{e}}Q(t)=d_{\text{e}}Q_{1}+d_{\text{e}}Q_{2}, \label{total=Exch-heat}%
\end{equation}
and have found
\begin{equation}
d\xi=d_{\text{e}}Q_{1}-d_{\text{e}}Q_{2}+2dE_{\text{in}} \label{internal-dxi}%
\end{equation}
using its general definition $d\xi(t)=dE_{1}(t)-dE_{2}(t)$. We thus see that
$d_{\text{i}}S(t)$ obtained by both realizations are the same as they must.
However, the realization $\Sigma_{\text{C}}(t)$ allows us to also identify
$d\xi$.\ 

Each exchange generates irreversible entropy following Eq. (\ref{diS_Sigma-s}%
). Using $d_{\text{e}}Q(t)=d_{\text{e}}Q_{1}(t)+d_{\text{e}}Q_{2}(t)$ in
$dQ(t)=T(t)dS(t)$ to determine $d_{\text{i}}Q(t)$, we find the generalization
of Eq. (\ref{Single-Interacting diS}):
\begin{equation}
d_{\text{i}}S(t)=\left[  \beta_{1}(t)-\beta_{2}(t)\right]  dQ_{\text{in}}(t)+%
{\textstyle\sum\nolimits_{l}}
[T_{0}\beta_{l}(t)-1]d_{\text{e}}S_{l}(t). \label{Composite-Interacting diS}%
\end{equation}
It is easy to see that the last term above gives nothing but the sum of the
irreversible entropy due to external exchange of heat by $\Sigma_{1}(t)$ and
$\Sigma_{2}(t)$ with $\widetilde{\Sigma}_{\text{h}}$:%
\begin{equation}
d_{\text{i}}S^{\text{Q}}(t)=d_{\text{i}}S_{1}^{\text{Q}}(t)+d_{\text{i}}%
S_{2}^{\text{Q}}(t), \label{diS_S1-S2}%
\end{equation}
where
\begin{equation}
d_{\text{i}}S_{l}^{\text{Q}}(t)=[T_{0}\beta_{l}(t)-1]d_{\text{e}}%
S_{l}(t),l=1,2 \label{diS_Sel}%
\end{equation}
is the external entropy exchange of $\Sigma_{l}(t)$ with $\widetilde{\Sigma
}_{\text{h}}$.

Thus, whether we treat $\Sigma$ as a system at temperature $T(t)$ or a
collection of $\Sigma_{1}(t)$ and $\Sigma_{2}(t)$ at temperatures $T_{1}(t)$
and $T_{2}(t)$, respectively, we obtain the same irreversibility. In other
words, $T(t)$ is a sensible thermodynamic temperature even in the presence of inhomogeneity.

\subsection{$\Sigma$ Interacting with $\widetilde{\Sigma}_{\text{h1}}$ and
$\widetilde{\Sigma}_{\text{h2}}$}

We now consider our composite $\Sigma$ in thermal contact with two distinct
and mutually noninteracting stochastic media $\widetilde{\Sigma}_{\text{h1}}$
and $\widetilde{\Sigma}_{\text{h2}}$ at temperatures $T_{01}$ and $T_{02}$. We
will again discuss the two different realizations as above.

(i) We first consider $\Sigma_{\text{B}}(t)$ at temperature $T(t)$, which
interacts with the two $\widetilde{\Sigma}_{\text{h}}$'s, and use the general
result in Eq. (\ref{diS-Media}). A simple calculation using $dS_{\text{B}}$
generalizes Eq. (\ref{Single-Interacting diS}) and yields%
\begin{subequations}
\begin{equation}
d_{\text{i}}S(t)=%
{\textstyle\sum\nolimits_{l}}
[\beta(t)-\beta_{0l}]d_{\text{e}}Q_{l}(t)+A^{\prime}(t)d\xi(t),
\label{diSb-two media}%
\end{equation}
since this reduces to that result when we set $\beta_{01}=\beta_{02}=\beta
_{0}$. As above, $d_{\text{i}}Q(t)=d_{\text{i}}W(t)=A(t)d\xi(t)$; see Eq.
(\ref{diS-Isolated-Composite}), which gives rise to the last term above. Thus,
setting $d_{\text{e}}Q_{l}(t)=0,l=1,2$ to make $\Sigma$ isolated,\ we retrieve
$d_{\text{i}}S(t)$ in Eq. (\ref{diS-Isolated-Composite}) as expected. The
first sum above gives the external entropy exchanges with the two heat media
as above.

(ii) We now consider $\Sigma_{\text{C}}$, and allow $\widetilde{\Sigma
}_{\text{h1}}$ to directly interact with $\Sigma_{1}(t)$ at temperature
$T_{1}(t)$ and $\widetilde{\Sigma}_{\text{h2}}$ to directly interact with
$\Sigma_{2}(t)$ at temperature $T_{2}(t)$. Using $dS_{\text{C}}$ generalizes
Eq. (\ref{Single-Interacting diS}) and yields%
\begin{equation}
d_{\text{i}}S(t)=%
{\textstyle\sum\nolimits_{l}}
[\beta_{l}(t)-\beta_{0l}]d_{\text{e}}Q_{l}(t)+\left[  \beta_{1}(t)-\beta
_{2}(t)\right]  dE_{\text{in}}(t). \label{diSc-two media}%
\end{equation}
Again using Eq. (\ref{beta1-beta2}) to express $\beta_{l}$, we can rewrite the
above $d_{\text{i}}S(t)$ as the $d_{\text{i}}S(t)$ in Eq.
(\ref{diSb-two media}) for $\Sigma_{\text{B}}$, and also find that $d\xi$ is
given by Eq. (\ref{internal-dxi}).

It should be emphasized that the determination of $d_{\text{i}}S(t)$ in Eqs.
(\ref{diSb-two media}-\ref{diSc-two media}) is valid for all cases of $\Sigma$
interacting with $\widetilde{\Sigma}_{\text{h1}}$ and $\widetilde{\Sigma
}_{\text{h2}}$ as we have not imposed any conditions on $T_{1}(t)$ and
$T_{2}(t)$ with respect to $T_{01}$ and $T_{02}$, respectively.\ Thus it is
very general. The derivation also applies to the NEQ stationary state, which
happens when $T_{1}(t)\rightarrow T_{01}$ and $T_{2}(t)\rightarrow T_{02}$.
For the stationary case, using Eq. (\ref{diSc-two media}), we have%
\end{subequations}
\begin{equation}
d_{\text{i}}S^{\text{st}}=\left[  \beta_{01}-\beta_{02}\right]  dE_{\text{in}%
}, \label{diS-stationary}%
\end{equation}
where all quantities on the right have their steady state values. Thus,
$d_{\text{i}}S^{\text{st}}$ is only determined by the stationary value of the
internal energy exchange $dE_{\text{in}}$. The reader can easily verify that
$d_{\text{i}}S(t)$ in Eqs. (\ref{diSb-two media}) also reduces to the above
result in the stationary limit.

From the above examples, we see that we can consider $\Sigma$\ in any of the
two realization $\Sigma_{\text{B}}$ and $\Sigma_{\text{C}}$ as we obtain the
same thermodynamics in that $d_{\text{i}}S(t)$ is identical. We emphasize this
important observation by summarizing it in the following conclusion.

\begin{conclusion}
If we consider $\Sigma(t)$ as a single system $\Sigma_{\text{B}}$ with an
uniform temperature $T(t)$ and with an internal variable $\xi(t)$, we do not
need to consider the energy transfer $dE_{\text{in}}(t)$ explicitly to obtain
$d_{\text{i}}S(t)$. If we consider $\Sigma(t)$ as a composite system
$\Sigma_{\text{C}}$ formed of $\Sigma_{1}(t)$ and $\Sigma_{2}(t)$ at their
specific temperatures, then we specifically need to consider the energy
transfer $dE_{\text{in}}(t)$ to obtain $d_{\text{i}}S(t)$ but no internal variable.
\end{conclusion}

This conclusion emphasizes the most important fact of the MNEQT that the
homogeneous thermodynamic temperature $T(t)$ of $\Sigma_{\text{B}}$\ can also
describe an inhomogeneous system $\Sigma_{\text{C}}$. This observation
justifies using the thermodynamic temperature $T(t)$ for treating $\Sigma(t)$
as a single system $\Sigma_{\text{B}}$, a black box, without any need to
consider the internal energy transfers.

The above discussion can be easily extended to also include inhomogeneities
such as two different work media $\widetilde{\Sigma}_{\text{w}}^{(1)}$ and
$\widetilde{\Sigma}_{\text{w}}^{(2)}$ corresponding to different pressures
$P_{01\text{ }}$and $P_{02}$. We will not do that here.

\subsection{$\Sigma$ Interacting with $\widetilde{\Sigma}_{\text{w}}$ and
$\widetilde{\Sigma}_{\text{h}}$}

In this case, $\Sigma$ is specified by two observables $E$ and $V$ so to
describe any inhomogeneity will require considering at least two subsystems
$\Sigma_{1}$ and $\Sigma_{2}$ specified by $E_{1},V_{1}$ and $E_{2,}V_{2}$,
respectively. From these four observables, we construct the following four
combinations%
\begin{align*}
E_{1}+E_{2}  &  =E,\xi_{\text{E}}=E_{1}-E_{2},\\
V_{1}+V_{2}  &  =V,\xi_{\text{V}}=V_{1}-V_{2},
\end{align*}
to express the entropy%
\[
S(E,V,\xi_{\text{E}},\xi_{\text{V}})=S_{1}(E_{1},V_{1})+S_{2}(V_{2},V_{2}).
\]
in terms of
\[
E_{1,2}=\frac{E\pm\xi_{\text{E}}}{2},V_{1,2}=\frac{V\pm\xi_{\text{V}}}{2}.
\]
Note that we have assumed that $\Sigma_{1}$ and $\Sigma_{2}$ are in EQ (no
internal variables for them). We now follow the procedure carried out in Sect.
\ref{Sec-Composite System} to identify thermodynamic temperature $T$, pressure
$P$, and affinities:%

\begin{equation}%
\begin{array}
[c]{c}%
\beta=\frac{(\beta_{1}+\beta_{2})}{2},\beta P=\frac{(\beta_{1}P_{1}+\beta
_{2}P_{2})}{2},\\
\beta A_{\text{E}}=\frac{(\beta_{1}-\beta_{2})}{2},\beta A_{\text{V}}%
=\frac{(\beta_{1}P_{1}-\beta_{2}P_{2})}{2}.
\end{array}
\label{System-Both Interactions}%
\end{equation}
All these quantities are SI-quantities and have the same values regardless of
whether $\Sigma$ is isolated or interacting. A more complicated
inhomogeneities will require more internal variables.

\begin{remark}
\label{Marker-Inhomogeneities}We now make an important remark about Eq.
(\ref{Irreversible EntropyGeneration-Complete}) that contains only a single
internal variable. From what is said above, it must include at least two
internal variables if $\Sigma$ contains inhomogeneity in both $E$ and~$V$. If
it contains inhomogeneity in only one variable, then and only then we will
have at least one internal variable. Thus, either we will have $\xi_{\text{E}%
}$ or $\xi_{\text{V}}$ as the case may be.
\begin{figure}
[ptb]
\begin{center}
\includegraphics[
height=2.1854in,
width=2.8444in
]%
{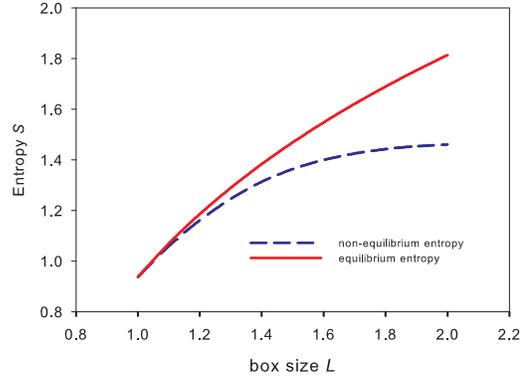}%
\caption{The calculated equilibrium (continuous) and nonequilibrium (broken)
entropies per particle for an ideal gas in a box as a function of the
expansion box length $L$. The nonequilibrium state is the result of a sudden
expansion from the initial state corresponding to $L=1$ and $T_{0}=4.$ The
energy of the gas remains constant in the sudden expansion. As expected, the
nonequilibrium entropy lies below the equilibrium entropy. In time, the former
will increase to the latter as the gas equilibrates. }%
\label{Fig.Entropy}%
\end{center}
\end{figure}

\end{remark}

\subsection{Free (Sudden) Expansion of the Box}

We consider the $1$-d ideal gas considered in Sect.
\ref{Sec-EntropyCalculation}. The box expands as a function of time, which
need not be quasi-static (extremely slow) so there is no reason to assume that
the gas remains in equilibrium after expansion. The entropy of the gas per
particle can be obtained by calculating $S(L,t)=-%
{\textstyle\sum\nolimits_{k}}
p_{k}(t)\ln p_{k}(t)$ for the particle under consideration so that the
irreversible entropy change $d_{\text{i}}S(t)$ will never be negative. The
discussion about the chemical reaction in Sect. \ref{Sec-EntropyCalculation}
shows that the change $d_{\text{i}}p_{k}(t)$ is caused by the transitions
between different eigenstates.

We prepare the gas in \emph{equilibrium} at some initial temperature
$T_{0\text{in}}$ in a box of length $L_{\text{in}}$, which we take to be
$L_{\text{in}}=1$. This is obtained by keeping the box in a medium of
temperature $T_{0\text{in}}$. The corresponding microstate probabilities
follow the Boltzmann law ($\beta_{\text{in}}\equiv1/T_{0\text{in}}$):%
\[
p_{k}^{\text{eq}}(\beta_{\text{in}},L_{\text{in}})=\exp(-\beta_{\text{in}%
}\varepsilon_{k}(L_{\text{in}}))/Z_{0}(\beta_{\text{in}},L_{\text{in}}),
\]
where $Z_{0}(\beta_{\text{in}},L_{\text{in}})\equiv%
{\textstyle\sum\nolimits_{k}}
\exp(-\beta_{\text{in}}\varepsilon_{k}(L_{\text{in}}))$ denotes the
equilibrium partition function; compare with Eq.
(\ref{microstate probability-EQ}). The initial energy per particle
$\varepsilon_{\text{in}}$ is obtained by replacing $p_{k}(\beta,L)$ by
$p_{k}^{\text{eq}}(\beta_{0},L_{\text{in}})$ in Eq.
(\ref{particle energy pressure}); the corresponding pressure is $P_{\text{in}%
}=2\varepsilon_{\text{in}}/L_{\text{in}}$. The equilibrium entropy can be
obtained by using the single particle probability $p_{k}^{\text{eq}}(\beta
_{0},L_{\text{in}})$ in Eq. (\ref{Gibbs_Formulation}).\ The initial
temperature $T_{0\text{in}}$ is taken to be $T_{0\text{in}}=4$, see Fig.
\ref{Fig.Entropy}, so that the initial energy $\varepsilon_{\text{in}}%
\approx2.786$.

We now consider NEQ states. For this, we isolate the box from its medium and
consider its free expansion as it expands suddenly from $L_{\text{in}}$ to a
new size $\ L>L_{\text{in}}$. Because of its isolation, its energy remains
$\varepsilon_{\text{in}}$ during this expansion. As the expansion is sudden,
the initial eigenfunctions $\psi_{l\text{in}}(x)$ for $L_{\text{in}}$ have no
time to change, but are no longer the eigenfunctions of the new size $L$; the
latter are given by $\psi_{k}(x)$ in Eq. (\ref{eigenfunctions}) for $L$.
However, $\psi_{l\text{in}}(x)$ can be expanded in terms of $\psi_{k}(x)$ as a
sum over $k$. The corresponding expansion coefficients $b_{kl}$ are easily
seen to be \cite{Bender}%
\[
b_{kl}(L,L_{\text{in}})=\frac{2l\alpha^{3/2}(-1)^{l}}{\pi(k^{2}-\alpha
^{2}l^{2})}\sin(\frac{k\pi}{\alpha}).
\]
Using $b_{kl}$, we can determine the probability $p_{k}(\beta_{0\text{in}%
},L,L_{\text{in}})$ for the $k$th microstate in the new box. We have checked
that the new probabilities add to $1$ and that the (average) energy after the
free expansion is equal to $\varepsilon_{\text{in}}$ to within our
computational accuracy. Thus, $\Delta\varepsilon=0$ in the sudden expansion.
This is consistent with the fact that the gas does no external work and that
no external heat is exchanged.

Despite this, the free expansion is spontaneous once the confining walls have
moved. Therefore, the (thermodynamic) entropy of the gas must increase in this
process in accordance with the second law. We use $p_{k}(\beta_{0\text{in}%
},L,L_{\text{in}})$\ to evaluate the nonequilibrium statistical entropy, which
is shown by the dashed curve in Fig. \ref{Fig.Entropy}. The significance of
this curve is as follows: Choose a particular value $L$ in this graph. Then,
the NEQ entropy for this $L$ is given by numerically evaluating the sum in Eq.
(\ref{Gibbs_Formulation}):
\[
S(\beta_{\text{in}},L,L_{\text{in}})=-%
{\textstyle\sum\nolimits_{k}}
\ p_{k}(\beta_{\text{in}},L,L_{\text{in}})\ln p_{k}(\beta_{\text{in}%
},L,L_{\text{in}}).\
\]
This is the entropy after the sudden expansion from the initial state at
$L_{\text{in}}=1$ and follows from the above quantum superposition principle.
Evidently, this entropy is higher than the initial equilibrium entropy
$S_{\text{eq}}(\beta_{\text{in}},L_{\text{in}})$. It is also obvious that this
entropy has a memory of the initial state at $L_{\text{in}}=1$ and
$T_{0\text{in}}=4$. Therefore, it does not represent the equilibrium entropy.
If we now wait at the new value of $L$, the isolated gas in the new box will
relax to approach its equilibrium state in which its nonequilibrium entropy
will gradually increase until it becomes equal to its value on the upper curve.

\section{1-d Tonks Gas:\ A simple Continuum Model}

A careful reader would have realized by this time that the proposed entropy
form in Eq. (\ref{Gibbs_Formulation}) is not at all the same as the standard
classical formulation of entropy, such as for the ideal gas, which can be
negative at low temperatures or at high pressures. The issue has been
discussed elsewhere \cite{Guj-Fedor} but with a very different perspective.
Here, we visit the same issue that allows us to investigate if and how the
entropy in continuum models is related to the proposed entropy in this work.
For this, we turn to a very simple continuum model in classical statistical
mechanics: the Tonks gas \cite{Tonks,Thompson}, which is an athermal model and
contains the ideal gas as a limiting case when the rod length $l$ vanishes. We
will simplify the discussion by considering the Tonks gas in one dimension.
The gas consists of $r$\ impenetrable rods, each of length $l$ lying along a
line of length $L$. We will assume $r$ to be fixed, but allow $l$ and $L$ to
change with the state of the system, such as its pressure.. The
configurational entropy per rod determined by the configurational partition
function is found to be \cite{Thompson}%
\begin{equation}
s_{\text{c}}=\ln[e(v-l)], \label{Tonks_S}%
\end{equation}
where $v$ is the "volume" available per rod $L/r$. Even though the above
result is derived for an equilibrium Tonks gas, it is easy to see that the
same result also applies for the gas in internal equilibrium. The only
difference is that the parameters in the model are also functions of internal
variables now.

The entropy vanishes when $v=l+1/e$ and becomes negative for all $v<l+1/e$.
Indeed, it diverges to $-\infty$ in the incompressible limit $v=l$. This is
contrary to the Boltzmann approach in which the entropy is determined by the
number of microstates (cf. Eq. (\ref{Boltzmann_S})) or the Gibbs approach (cf.
Eq. (\ref{Gibbs_Formulation})) and can never be negative. Can we reconcile the
contradiction between the continuum entropy and the current statistical formulation?

We now demonstrate that the above entropy for the Tonks gas is derivable from
the current statistical approach under some approximation, to be noted below,
by first considering a lattice model for the Tonks gas and then taking its
continuum limit. It is in the lattice model can we determine the number of
microstates. In a continuum, this number is always \emph{unbounded }(see below
also). For this we consider a $1$-d lattice $\Lambda_{\text{f}}$ with
$N_{\text{f}}$ sites; the lattice spacing, the distance between two
consecutive sites, is given by $\delta$. We take $N_{\text{f}}>>1$ so that
$L_{\text{f}}=(N_{\text{f}}-1)\delta\approx N_{\text{f}}\delta$ is the length
of the the lattice $\Lambda_{\text{f}}$. We randomly select $r$ sites out of
$N_{\text{f}}$. The number of ways, which then represents the number of
configurational microstates, is given by%
\begin{equation}
W_{\text{c}}=N_{\text{f}}!/r!(N_{\text{f}}-r)!. \label{Tonks_Microstates}%
\end{equation}
After the choice is made, we replace each selected site by $\lambda+1$
consecutive sites, each site representing an atoms in a rod, to give rise to a
rod of length $l\equiv\lambda\delta$. It is clear that $\delta$\ also changes
with the state of the system. The number of sites in the resulting lattice
$\Lambda$ is
\[
N=N_{\text{f}}+r\lambda
\]
so that the length of $\Lambda$ is given by $L=(N-1)\delta\approx N\delta$
since $N>>1$. We introduce various densities $\varphi_{\text{f}}%
=r/N_{\text{f}}$, $\rho_{\text{f}}=r/N_{\text{f}}\delta\approx r/L_{\text{f}}$
and $\rho=r/N\delta\approx r/L$. A simple calculation shows that $S=\ln
W_{\text{c}}$ is given by%
\[
S=-N_{\text{f}}[\rho_{\text{f}}\delta\ln\rho_{\text{f}}\delta+\ln
(1-\rho_{\text{f}}\delta)-\rho_{\text{f}}\delta\ln(1-\rho_{\text{f}}\delta)].
\]
This result can also be obtained by taking the athermal entropy for a
polydisperse polymer solution a Bethe lattice \cite{Gujrati-Monodisperse} by
setting the coordination number $q$ to be $q=2$. We now take the continuum
limit $\delta\rightarrow0$ \ for \emph{fixed} $\rho_{\text{f}}$ and $\rho$,
that is \emph{fixed} $L_{\text{f}}$ and $\rho L$, respectively. In this limit,
$\ln(1-\rho_{\text{f}}\delta)\approx-\rho_{\text{f}}\delta$, and
$\rho_{\text{f}}\delta\ln(1-\rho_{\text{f}}\delta)\approx-(\rho_{\text{f}%
}\delta)^{2}$. Use of these limits in $S$ yields%
\begin{equation}
S=-r\ln(e/\rho_{\text{f}}\delta)\rightarrow\infty. \label{Cont_S}%
\end{equation}
The continuum limit of the entropy from the Boltzmann approach has resulted in
a diverging entropy regardless of the value of $\rho_{\text{f}}$
\cite{Guj-Fedor}, a well known result. By introducing an arbitrary
\emph{constant} $a$ with the dimension of length, we can rewrite $S$ as%
\begin{equation}
S/r=-\ln(e/\rho_{\text{f}}a)+\ln(\delta/a), \label{Cont_Manipulation}%
\end{equation}
in which the first term remains finite in the continuum limit, and the second
term contains the divergence. The diverging part, although explicitly
independent of $\rho_{\text{f}}$, still depends on the state of the gas
through $\delta$, and cannot be treated as a constant unless we assume
$\delta$\ to be independent of the state of the gas. It is a common practice
to approximate the lattice spacing $\delta$ as a constant. In that case, the
diverging term represents a constant that can be subtracted from $S/r$.
Recognizing that $1/\rho_{\text{f}}=v-l$, we see that the first term in Eq.
(\ref{Cont_Manipulation}) is nothing but the entropy of the Tonks gas in Eq.
(\ref{Tonks_S}) for the arbitrary constant $a=1$. However, this equivalence
only occurs in the state independent constant-$\delta$ approximation.

As the second term above has been discarded, the continuum entropy
$s_{\text{c}}$ also has \emph{no} simple relationship with the number ($\geq
1$) of microstates in the continuum limit, which means that the continuum
entropy cannot be identified as the Boltzmann entropy in Eq.
(\ref{S_Boltzmann}). To see this more clearly, let us focus on the centers of
mass of each rod, which represent one of the $r$ sites that were selected in
$\Lambda_{\text{f}}$. Each of the $k$ sites $x_{k}$, $k=1,2,\cdots,r$, is free
to move over $L_{\text{f}}$. The \emph{adimensional} volume $\left\vert
\Gamma_{\text{f}}\right\vert $, also called the probability and denoted by $Z$
by Boltzmann,\cite{Note-Boltzmann,Lebowitz} of the corresponding phase space
$\Gamma_{\text{f}}$ is $L_{\text{f}}^{r}/a^{r}$. However, contrary to the
conventional wisdom \cite{Lebowitz}, $\ln\left\vert \Gamma_{\text{f}%
}\right\vert $ does not yield $s_{\text{c}}$. The correct expression is given
by the Gibbs-modified adimensional volume $\left\vert \Gamma_{\text{f}%
}\right\vert /r!$, i.e.
\[
\frac{1}{r!a^{r}}L_{\text{f}}^{r}.
\]
The presence of $r!$ is required to restrict the volume due to
indistinguishability of the rods \`{a} la Gibbs. For large $r$, this quantity
correctly gives the entropy $s_{\text{c}}$. However, this quantity is not only
not an integer, it also cannot be \emph{always} larger than or equal to unity,
as noted above.

\section{Jaynes Revisited\label{Marker_Jaynes}}

Boltzmann \cite{Boltzmann} provides the following alternative expression of
the entropy \cite{Cohen,Boltzmann} in terms of a single particle probability
$p_{i}^{(1)}$ for the particle to be in the $i$th state:%
\begin{equation}
S_{\text{B}}^{(1)}=-N%
{\textstyle\sum\nolimits_{i}}
p_{i}^{(1)}\ln p_{i}^{(1)}, \label{Boltzmann_S_1}%
\end{equation}
not to be confused with that in Eq. (\ref{Boltzmann_S}). Boltzmann is only
interested in the maximum entropy, which occurs when all states are equally
probable. In this case,
\[
S_{\text{B, max}}^{(1)}=N\ln w
\]
where $w$ is the number of possible states of a single particle in the gas. In
general, particles are not independent due to interactions and number of
possible states $W<w^{N}$. Accordingly, maximum Gibbs entropy $S_{\text{ max}%
}$ per particle is \emph{less} than the corresponding equiprobable Boltzmann
entropy $S_{\text{B, max}}^{(1)}$. However, Jaynes \cite[see Eq. (5)
there]{Jaynes} gives a much stronger results:%
\[
S<S_{\text{B}}^{(1)}.
\]
The equality occurs only if there are no interactions between the particles,
as we have asserted above.\ \ \ 

\section{Summary and Discussion\label{marker_Summary}}

Recognizing that there does not exists a thermodynamic definition of the
classical Clausius entropy for a NEQ state $\mathcal{M}$ of a body
$\Sigma_{\text{b}}$, we have proposed a way to identify it by choosing a large
enough state space $\mathfrak{S}_{\mathbf{Z}^{\prime}\text{ }}$in which the
entropy becomes a state function $S(\mathbf{Z}^{\prime})$ so that
$\mathcal{M}$ becomes uniquely defined in this space. A unique state in an
extended state space has been called an IEQ state $\mathcal{M}_{\text{ieq}}$.
However, $\mathcal{M}_{\text{ieq}}$ is no longer unique, which we denote by
$\mathcal{M}_{\text{nieq}}$, in a smaller subspace $\mathfrak{S}_{\mathbf{Z}%
}\subset$ $\mathfrak{S}_{\mathbf{Z}^{\prime}}$, where its entropy
$S(\mathbf{Z},t)$ is also not a state function. This is discussed in Sect.
\ref{Sec-M_nieq}. The Gibbs fundamental relation in Eq.
(\ref{GibbsFR-S-NIEQ-prime}) for $S(\mathbf{Z}^{\prime})$ reduces to an
effective Gibbs fundamental relation in Eq. (\ref{GibbsFR-S-NIEQ}), in which
the partial time derivative $\partial S/\partial t$ is found to be related to
the hidden entropy generation%
\[
d_{\text{i}}S^{\text{hid}}(t)\equiv(\partial S/\partial t)dt
\]
due to the hidden internal variable $\boldsymbol{\xi}^{\prime}$. This entropy
generation is in addition to $d_{\text{i}}S$ in $\mathfrak{S}_{\mathbf{Z}}$.
The sum of the two is the net entropy generation in $\mathfrak{S}%
_{\mathbf{Z}^{\prime}}$. Such an identification is the reason for Proposition
\ref{Proposition-General-MNEQT}, according to which one can always identify a
state space $\mathfrak{S}_{\mathbf{Z}^{\prime}}$ in which an arbitrary
$\mathcal{M}$ becomes $\mathcal{M}_{\text{ieq}}$. The rational for this is the
hierarchy of equilibration times $\tau_{i}$ of internal variables and its
interplay with $\tau_{\text{obs}}$\ that has been recently discussed by us
\cite{Gujrati-Hierarchy}, and which is briefly reviewed in Sect.
\ref{Sec-Choice-S_Z}. Thus, one can, in principle, identify $\mathfrak{S}%
_{\mathbf{Z}^{\prime}}$ in which any $\mathcal{M}$ can be uniquely specified.
This then leads to a state function $S(\mathbf{Z}^{\prime})$. However, for the
purpose of computation, one needs to choose a small enough subspace
$\mathfrak{S}_{\mathbf{Z}}\subset$ $\mathfrak{S}_{\mathbf{Z}^{\prime}}$ in
which the same $\mathcal{M}$ will become nonunique $\mathcal{M}_{\text{nieq}}%
$, with an entropy $S(\mathbf{Z},t)$.

Unfortunately, having a Gibbs fundamental relation does not allow for a way to
experimentally "measure" $S(\mathbf{Z}^{\prime})$ or $S(\mathbf{Z},t)$. This
can only be done for $\mathcal{M}_{\text{eq}}$. To overcome this limitation,
we propose a statistical definition $\mathcal{S}$ of the entropy. By giving a
first principles statistical formulation of a NEQ\ $S$ for $\Sigma_{\text{b}}$
in terms of microstate probabilities, we have attempted to fill in the gap. We
use a formal approach (frequentist interpretation of probability) by extending
the EQ ensemble of Gibbs in $\mathfrak{S}_{\mathbf{X}}$ to a NEQ ensemble,
which is nothing but a large number $\mathcal{N}$ of samples of the
thermodynamic system under consideration in the enlarged state space. We refer
to the ensemble as a sample space. The formal approach enables us to evaluate
the combinatorics for a given set of microstate probabilities. The resulting
statistical entropy is independent of the number of samples and depends only
on the probabilities as is seen from Eqs. (\ref{Gibbs_Formulation}) and
(\ref{S_Component}). Thus, the use of a large number of samples is merely a
formality and is not required in practice as we can use Eq.
(\ref{microstate probability}) for any computation. The resulting statistical
entropy $\mathcal{S}$ is shown to be identical to $S(\mathbf{Z}^{\prime})$ for
$\mathcal{M}_{\text{ieq}}$\ or $S(\mathbf{Z},t)$ for $\mathcal{M}%
_{\text{nieq}}$, which establishes their equivalence for any arbitrary state,
EQ or not, and generalizes our previous result
\cite{Gujrati-Entropy1,Gujrati-Entropy2} that only shows their equivalence
($\mathcal{S}(\mathbf{X})=S(\mathbf{X})$) for $\mathcal{M}_{\text{eq}}$\ or
($\mathcal{S}(\mathbf{Z})=S(\mathbf{Z})$)$\ $for $\mathcal{M}_{\text{ieq}}$.
Our demonstration goes beyond the standard practice to use the classical
nonequilibrium thermodynamics \cite{deGroot} or its variant to calculate NEQ
entropy. In this approach, one treats the entropy at the local level as a
state function without any internal variables. In contrast, our approach
allows us to theoretically "measure" the classical entropy $S(\mathbf{Z}%
^{\prime})$ for $\mathcal{M}_{\text{ieq}}$\ or $S(\mathbf{Z},t)$ for
$\mathcal{M}_{\text{nieq}}$ by knowing the probabilities $p_{k}$, and use Eq.
(\ref{Gibbs_Formulation}) to evaluate $S=\mathcal{S}$.

The choice of the appropriate state space $\mathfrak{S}_{\mathbf{Z}}$ for
$\mathcal{M}$ to become $\mathcal{M}_{\text{ieq}}$ is determined by the
observation time $\tau_{\text{obs}}$ as discussed in Sect.
\ref{Sec-Choice-S_Z}. We have given several examples to show how to identify
the internal variables in a unique manner for some $\Sigma_{\text{b}}$. Once
$\mathfrak{S}_{\mathbf{Z}}$ has been identified, we can identify all the
fields that have thermodynamic significance. Among these is the temperature
$T$, which plays the role of a global temperature over $\Sigma_{\text{b}}$. In
addition, the internal variable $\boldsymbol{\xi}$ captures the entire
internally produced irreversibility so we can treat $\Sigma_{\text{b}}$ as a
black box $\Sigma_{\text{B}}$ as discussed in Sect. \ref{Sec-Applications} so
that we do not need to know anything about its internal structure. We have
shown that we obtain the same description by not using $\boldsymbol{\xi}$
\ and treating\ $\Sigma_{\text{b}}$ as a composite box $\Sigma_{\text{C}}$
requiring the knowledge of its internal structure. Both descriptions are,
therefore, equivalent but we believe the extended state space description to
be less tedious and more revealing.

Some readers may think that our statistical formulation is no different than
that used in the information theory \cite{Wiener,Shannon}. We disagree. We
refer the reader to an excellent overview \cite{Bawden} on this topic. As
pointed out by Jaynes \cite{Jaynes0}, information entropy can be related to
our EQ statistical mechanical entropy. Our concern is with NEQ\ $\mathcal{S}$
and the information entropy. Here, we limit the discussion to the following
relevant issues that highlight their differences. For one, there is no concept
of the temperature, and internal variable $\boldsymbol{\xi}$ in the latter
entropy. Because of this, our approach allows us to consider multi-level
descriptions so that we can consider several different entropies
$S(\mathbf{Z}_{n},t),S(\mathbf{Z}_{n}),n\geq1,S(\mathbf{X},t)$ and
$S(\mathbf{X})$ satisfying the inequalities in Eq.
(\ref{Entropy_Inequalities0}). The information theory can only deal with two
levels of entropies. There is also no concept of relaxation and dynamics and
the concept of $\tau_{\text{obs}}$ so there is also no concept of component
confinement in time to explain a residual entropy in the latter. It is also
not clear if there is an analog of $d_{\text{e}}p_{k}$\ and $d_{\text{i}}%
p_{k}$\ and of the second law.

For an isolated system in internal equilibrium ($p_{k}=p$ for $\forall m_{k}%
$), just a single sample will suffice to determine the entropy as samples are
\emph{unbiased}. The entropy in this case is no different than the "entropy"
$-\ln p$ of a single sample \cite{Gujrati-Symmetry,Lebowitz}:%
\[
S(t)=(-p\ln p)W=-\ln p=\ln W,
\]
where $W$ represents $W(\mathbf{Z})$ or $W(\mathbf{X})$.

Changes in microstate probabilities result in changes in the entropy. There
are two ways probabilities can change within an isolated system, both of them
being irreversible in nature. One cause of changes is due to the quantum
nature as seen in the sudden expansion of the box. Here, the parameter
$\lambda$ ($=L$) changes non-adiabatically and creates irreversibility. The
resulting irreversible change in the entropy for the $1$-$d$ gas has been
calculated and shown by the lower curve in Fig. \ref{Fig.Entropy}. The other
cause of probability changes is due to the "chemical reaction" going on among
the microstates that brings about equilibration in the system. The
corresponding irreversible rise in the entropy for the gas is shown by the
difference between the two curves in Fig. \ref{Fig.Entropy}. The interaction
of a body with its medium can also result in the changes in microstate
probabilities, and has been considered elsewhere \cite{Gujrati-Heat-Work}.

By considering a lattice model of the Tonks gas, for which the entropy remains
nonnegative, we show how its continuum limit results in a negative entropy.
This strongly suggests that negative entropies arise from a continuum model
such as $S_{f}$ in Eq. (\ref{Sf_S}), which in some cases can diverge to
$-\infty$, even though the statistical entropy remains finite and positive. We
should also point out that the definition of $S_{f}$ in Eq. (\ref{Sf_S}) makes
no sense as $f$ has the dimension of inverse volume in the phase space so $\ln
f$ has no meaning.\ We suggest that our statistical Gibbs formulation can be
applied to any nonequilibrium state under any condition.


\begin{thebibliography}{99}
\bibitem {Note-Boltzmann}The number of combinations in Eq. (35) on p. 56 in
Boltzmann \cite{Boltzmann0} is denoted by $Z$, but it is not the number of
microstates. The two become the same only when $Z$ is maximized as discussed
on p. 58.
\end{thebibliography}
\end{document}